\documentclass[11pt,oneside]{article}

\usepackage[paper=a4paper, total={16cm,23cm}]{geometry}
\usepackage[T1]{fontenc}
\usepackage{setspace} 
\usepackage[hang]{footmisc} 
\usepackage{titling} 
\usepackage{tocloft} 
\usepackage{parskip} 

\usepackage{amsfonts}
\usepackage{amsmath}
\usepackage{amssymb}
\usepackage{marvosym} 
\usepackage{physics}
\usepackage{mathtools}
\usepackage{upgreek} 
\numberwithin{equation}{section} 

\usepackage{xcolor}
\definecolor{dark-red}{rgb}{0.50,0.12,0.12} 

\usepackage{hyperref} 
\hypersetup{colorlinks=true,allcolors=dark-red,linktoc=all}

\usepackage{tikz}
\usepackage[subrefformat=parens]{subcaption}
\usepackage{pgfplots}
\pgfplotsset{compat=newest}

\def \cC {\mathcal{C}}
\def \cF {\mathcal{F}}
\def \cH {\mathcal{H}}
\def \cM {\mathcal{M}}
\def \cO {\mathcal{O}}
\def \cT {\mathcal{T}}
\def \cV {\mathcal{V}}
\def \cZ {\mathcal{Z}}

\def \sS {\mathsf{S}}
\def \sV {\mathsf{V}}

\def \rD {\mathrm{D}}
\def \rL {\mathrm{L}}

\def \rS {\mathrm{S}}
\def \rV {\mathrm{V}}
\def \rY {\mathrm{Y}}
\def \rmc {\mathrm{c}}
\def \rms {\mathrm{s}}

\def \bbC {\mathbb{C}}
\def \bbH {\mathbb{H}}
\def \bbR {\mathbb{R}}
\def \bbS {\mathbb{S}}
\def \bbZ {\mathbb{Z}}

\newcommand{\ep}{\mathrm{e}}
\newcommand{\ic}{\mathrm{i}}
\newcommand{\ua}{\upalpha}
\newcommand{\ub}{\upbeta}
\newcommand{\ug}{\upgamma}
\newcommand{\MCG}{\mathrm{MCG}}
\newcommand{\SL}{\mathrm{SL}}
\newcommand{\PSL}{\mathrm{PSL}}
\newcommand{\WP}{\mathrm{WP}}
\newcommand{\diff}{\mathrm{d}}
\newcommand{\Diff}{\rD}

\DeclareMathOperator{\arccosh}{arccosh}

\makeatletter
\newcommand{\defeq}{\mathrel{\rlap{\raisebox{0.3ex}{$\m@th\cdot$}}\raisebox{-0.3ex}{$\m@th\cdot$}}=}
\makeatother

\newcommand{\conj}[1]{\mkern 1.3mu\overline{\mkern-1.3mu#1\mkern-1.3mu}\mkern 1.3mu}

\makeatletter
\newcommand{\subalign}[1]{
  \vcenter{
    \Let@ \restore@math@cr \default@tag
    \baselineskip\fontdimen10 \scriptfont\tw@
    \advance\baselineskip\fontdimen12 \scriptfont\tw@
    \lineskip\thr@@\fontdimen8 \scriptfont\thr@@
    \lineskiplimit\lineskip
    \ialign{\hfil$\m@th\scriptstyle##$&$\m@th\scriptstyle{}##$\hfil\crcr
      #1\crcr
    }
  }
}
\makeatother

\begin{document}
\begin{titlingpage}
    \vspace*{3em}
    \onehalfspacing
    \begin{center}
        {\LARGE {Liouville} theory and the {Weil}--{Petersson} geometry of moduli space: bordered, conic, and higher genus surfaces}
    \end{center}
    \singlespacing
    \vspace*{2em}
    \begin{center}
        \textbf{
        Kale Colville,\textsuperscript{\Saturn}
        Sarah M. Harrison,\textsuperscript{\Jupiter,\Saturn,\Uranus}
        Alexander Maloney,\textsuperscript{\Saturn}
        and Keivan Namjou\textsuperscript{\Saturn}
        }
    \end{center}
    \vspace*{1em}
    \begin{center}
        \textsl{
        \Saturn\ Department of Physics, McGill University \\
        Montr\'{e}al, QC, Canada \\
        \Jupiter\ Department of Physics and Department of Mathematics, Northeastern University \\
        Boston, MA, USA \\
        \Uranus\ Department of Mathematics and Statistics, McGill University \\
        Montr\'{e}al, QC, Canada \\[\baselineskip]
        }
        \href{mailto:kale.colville@mail.mcgill.ca}{\small kale.colville@mail.mcgill.ca},
        \href{mailto:s.harrison@northeastern.edu}{\small s.harrison@northeastern.edu} \\
        \href{mailto:alex.maloney@mcgill.ca}{\small alex.maloney@mcgill.ca},
        \href{mailto:keivan.namjou@mail.mcgill.ca}{\small keivan.namjou@mail.mcgill.ca}
    \end{center}
    \vspace*{3em}
    \begin{abstract}
        Two-dimensional conformal field theory is a powerful tool to understand the geometry of surfaces.  Here, we study Liouville conformal field theory in the classical (large central charge) limit, where it encodes the geometry of the moduli space of Riemann surfaces.  Generalizing previous work, we employ this to study moduli spaces of higher genus surfaces, surfaces with boundaries, and surfaces with cone points.  In each case, the knowledge of classical conformal blocks provides an extremely efficient approximation to the Weil--Petersson metric on moduli space. We find detailed agreement with analytic results for volumes and geodesic lengths on moduli space.
    \end{abstract}
\end{titlingpage}
\tableofcontents

\section{Introduction}
Conformal field theory is a powerful tool in the study of geometry, with applications ranging from mirror symmetry to spectral theory and quantum chaos. Our focus in this paper is on 
Liouville theory \cite{Polyakov:1981rd}, which provides a natural means to study the Weil--Petersson geometry of the moduli space of Riemann surfaces.\footnote{See, e.g., \cite{Takhtajan:1993vt,Matone:1993tj,Takhtajan:1994vt} for early discussions of the relation between Liouville theory and the geometry of surfaces.} In the semiclassical limit, the Liouville path integral computes the K\"{a}hler potential, giving access to the metric and associated geometric quantities. Liouville conformal blocks can be computed efficiently \cite{Zamolodchikov:1987}, so this allows for detailed computations with surprising numerical accuracy. The study of the geometry of hyperbolic surfaces using Liouville theory was started in \cite{Hadasz:2005gk}, where the authors studied the hyperbolic metric on the four-puncture sphere using the conformal blocks. The solution to the saddle-point equations of classical Liouville theory can be used to compute the classical Liouville action, which is the K\"{a}hler potential on moduli space. This technique was applied in \cite{Harrison:2022frl} to efficiently compute the metric and Laplacian eigenvalues on the moduli space of the four-punctured sphere.

In the present work, we extend this to general Riemann surfaces. Despite increased complexity, the conformal field theory approach remains viable. As illustrative examples, we focus on the moduli spaces of spheres with cone points, of genus two surfaces with $\bbZ_3$ replica symmetry, and bordered Riemann surfaces. In all cases, we numerically compute moduli space volumes and geodesic lengths and find excellent agreement with analytic results in the special cases (where the surfaces have additional discrete symmetries) where exact results are available. We initiate a study of the spectra of Weil--Petersson Laplacians, which are generally only computable numerically.

Our results pave the way for new investigations. With this technique, detailed studies of Weil--Petersson geometry appear feasible. We anticipate applications in geometry, number theory, and physics. For instance, the eigenvalues of the Laplacian exhibit quantum chaotic behavior, and in some cases (see \cite{Harrison:2022frl}) agree with random matrix theory statistics. It would be fascinating to continue the study of these eigenvalues in detail in the examples we present in this paper. This also suggests intriguing connections to dual gravitational systems.

More broadly, the ability to numerically access geometrical information opens new possibilities to test conjectures, formulate hypotheses, and explore novel properties of these moduli spaces. Analytical calculations are typically restricted to low genus or certain limits, so numerical computation provides a vital complementary tool. Our method may uncover new structures in the Weil--Petersson geometry. The excellent agreement with analytical results validates the power of the approach and highlights its promise. We anticipate a broad impact spanning geometry, number theory, quantum gravity, black holes, and more.

On the physics side, the quantization of moduli space is closely related to quantum gravity in three dimensions. The Liouville action is in many cases closely related to (if not exactly equal to) the Einstein action of an associated three-manifold. Indeed, when spacetime has the topology of a surface $\Sigma$ times time, the constraint equations of 3d gravity with a negative cosmological constant can be solved exactly: the resulting phase space is the cotangent bundle of the moduli space of Riemann surfaces. Thus, our techniques offer insight into the structure of quantum gravity, and potentially even a possible statistical mechanical origin for black hole entropy \cite{Maloney:2015ina}.

The outline of the rest of the paper is as follows. In \S\ref{s:surfaces}, we review salient mathematical aspects of moduli spaces of hyperbolic Riemann surfaces, before reviewing Liouville theory in \S\ref{s:liouville}. We then present our method for numerically solving the saddle-point equations for arbitrary hyperbolic Riemann surface in \S\ref{s:metric}. As mentioned above, we present a number of examples in \S\ref{s:examples} to test this method---moduli spaces of spheres with cone points, of genus two surfaces with $\bbZ_3$ replica symmetry, and bordered Riemann surfaces---and compute both volumes of moduli spaces and geodesic lengths in each of these examples.

Finally, we relegate additional details to some appendices. In \S\ref{a:twistblocks}, we present our calculation of conformal blocks on a $\bbZ_3$-symmetric genus two surface.  In \S\ref{a:semiclassics}, we present the derivation of the semiclassical limits of some of the special functions used in the main body of the text. We give the values of the lowest 24 eigenvalues of the Weil--Petersson laplacian on the moduli space of $\bbZ_3$-symmetric genus two surfaces, computed numerically in \S\ref{a:eigs}.

\section{Moduli spaces of hyperbolic Riemann surfaces}\label{s:surfaces}
In this section, we briefly review some aspects of hyperbolic Riemann surfaces and their moduli spaces (\S\ref{s:hypergeom} and \S\ref{s:modspaces}), including cases where the surface has boundaries, punctures, and/or cone points. We also review certain classes of Riemann surfaces without boundary, which can be written as algebraic curves referred to as ``replica surfaces'' because of their connection with the computation of R\'{e}nyi entropies (\S\ref{s:algcurves}). Finally, we discuss analytic results for the volumes of moduli spaces of hyperbolic Riemann surfaces (\S\ref{s:volumes}).

\subsection{Hyperbolic Riemann surfaces}\label{s:hypergeom}
Consider a Riemann surface $\Sigma_{g,n}$ of genus $g$ with $n$ boundary components. The topological properties of this surface are fixed by its Euler characteristic, $\chi(\Sigma_{g,n}) = 2 - 2g - n$. The Gauss--Bonnet theorem relates the topology of the surface to its curvature,
\begin{equation}
    \int_{\Sigma_{g,n}} \diff A \, K + \int_{\partial \Sigma_{g,n}} \diff s \, k = 2\pi \, \chi(\Sigma_{g,n}),
\end{equation}
where $K$ is the Gaussian curvature\footnote{In two dimensions, the Ricci scalar is twice the Gaussian curvature.} of the surface $\Sigma_{g,n}$, $k$ is the geodesic curvature of its boundary $\partial \Sigma_{g,n}$, $\diff A$ is the area element on $\Sigma_{g,n}$, and $\diff s$ is the line element on $\partial \Sigma_{g,n}$.

On a surface $\Sigma_{g,n}(\boldsymbol{L})$ whose boundary components are $n$ geodesics of lengths $\boldsymbol{L}=(L_1, \ldots, L_n)$, the geodesic curvature along the boundary vanishes. Hence, for a metric with constant negative curvature to exist on the surface, the Euler characteristic must be negative. A hyperbolic Riemann surface is a Riemann surface that admits a constant negative curvature metric which is unique up to conformal transformations. In this paper, we will consider more general hyperbolic surfaces that we still denote as $\Sigma_{g,n}(\boldsymbol{L})$, where the boundary components can be punctures, i.e., $L_i = 0$, or cone points, which can be found by taking $L_i = - \ic \theta_i$ to be purely imaginary for a real defect angle $\theta_i$.

The uniformization theorem for hyperbolic Riemann surfaces ensures that for every hyperbolic surface $\Sigma$ there exists a Fuchsian group\footnote{A Fuchsian group $\Gamma$ is a discrete subgroup of the isometry group $\PSL(2,\bbR)$ of the upper half-plane $\bbH$.} $\Gamma \subset \PSL(2,\bbR)$ that is unique up to conjugation in $\PSL(2,\bbR)$ such that
\begin{equation}
    \Sigma \cong \bbH_2 / \Gamma.
\end{equation}
Here $\bbH_2 := \{z \in \bbC \mid \Im z > 0 \}$ is the upper half-plane with line element
\begin{equation}
    \diff s^2 = \frac{\diff z \diff \conj{z}}{\qty(\Im z)^2},
\end{equation}
 corresponding to the Poincar\'{e} metric. The isometry group of $\bbH_2$ is $\PSL(2,\bbR)$, whose elements act on the space as M\"{o}bius transformations
\begin{equation}
    \gamma: z \mapsto \gamma \cdot z = \frac{a z + b}{c z + d}, \qquad \gamma = \begin{pmatrix}
        a & b \\
        c & d
    \end{pmatrix} \in \PSL(2,\bbR).
\end{equation}
Since $\bbH_2$ is simply connected, the Fuchsian group $\Gamma$ is isomorphic to the fundamental group $\pi_1(\Sigma)$ of the surface $\Sigma \cong \bbH_2 / \Gamma$ obtained from the quotient construction.

The Fuchsian group $\Gamma_{\sigma}$ with signature $\sigma = (g,n; \nu_1, \dots, \nu_n)$ for a surface $\Sigma_{g,n}(\boldsymbol{L})$ with genus $g$ and $n$ singularities can be represented as
\begin{equation}
    \Gamma_\sigma = \langle \ua_1, \ub_1, \dots, \ua_g, \ub_g; \ug_1, \dots, \ug_n : \textstyle{\prod}_{i} [\ua_i, \ub_i] \textstyle{\prod}_i \ug_i = \ug_1^{\nu_1} = \cdots = \ug_n^{\nu_n} = 1 \rangle.
\end{equation}
Here $\ua_j$ and $\ub_j$ represent the usual $\ua$ and $\ub$ cycles in a higher genus Riemann surface, and $\ug_j$ represent the non-contractible cycles around the $n$ singularities. Each homotopy class on the Riemann surface has a geodesic\footnote{defined with respect to the hyperbolic metric on the surface.} representative whose length $d$ can be obtained by the following relation:
\begin{eqnarray}
    \cosh \frac{d}{2} = \frac{1}{2} \abs{\Tr h},
\end{eqnarray}
where $h \in \Gamma_\sigma$ is the corresponding element in the Fuchsian group that can be a combination of different cycles,
\begin{equation}
    h = h_{i_1} \cdots h_{i_k}, \qquad h_{i_j} \in \{\ua_1,\dots,\ua_g,\ub_1,\dots,\ub_g,\ug_1,\dots,\ug_n\}.
\end{equation}

The generator $\ug_j$ has the following properties depending on whether it leads to a geodesic boundary, puncture, or conical deficit:
\begin{itemize}
\item For hyperbolic elements $\abs{\Tr \ug_j} > 2$, and this element corresponds to a geodesic boundary of length $L_j$ given by,
\begin{equation}
    \cosh \frac{L_{j}}{2} = \frac{1}{2} \abs{\Tr \ug_j }.
\end{equation}
\item Parabolic elements $\ug_j$ satisfy $\abs{\Tr \ug_j} = 2$ and have $\nu_j = \infty$. For example, the Fuchsian group corresponding to an $n$-punctured sphere, i.e., with all elements parabolic, is
\begin{equation}
    \Gamma_{(0,n;\infty)} = \langle \ug_1, \dots, \ug_n : \textstyle{\prod}_i \ug_i = 1 \rangle.
\end{equation}
In this example, for any pairs of punctures $z_j$ and $z_k$ there is a unique closed geodesic with length $d_{jk} \defeq d(z_j,z_k)$ that encloses $z_j$ and $z_k$ and none of the other punctures, where
\begin{equation}
    \cosh \frac{d_{jk}}{2} = \frac{1}{2} \abs{\Tr \ug_j \ug_k}.
\end{equation}
\item Finally, elliptic elements correspond to $\abs{\Tr \ug_j} < 2$ and conical singularities with defect angle $\theta_j = \frac{2\pi}{\nu_j}$ where
\begin{equation}
    \cos \frac{\theta_j}{2} = \frac{1}{2} \abs{\Tr \ug_j }.
\end{equation}
\end{itemize}

\subsection{Moduli spaces}\label{s:modspaces}
The moduli space $\cM_{g,n}(\boldsymbol{L})$ of a Riemann surface $\Sigma_{g,n}(\boldsymbol{L})$ is defined as
\begin{equation}
    \cM_{g,n}(\boldsymbol{L}) = \left. \qty{ \qty(\sS, B_1, \dots, B_n) \, \middle| \,
    \begin{array}{l}
    \sS \text{ is hyperbolic surface of genus $g$ } \\
    \text{with $n$ boundaries $B_1, \dots, B_n$}
    \end{array}} \middle/ \sim \right.,
\end{equation}
where the boundary $B_j$ has geodesic length $L_j$, and $\qty(\sS,B_1, \dots, B_n) \sim \qty(\sS',B'_1, \dots, B'_n)$ if and only if there exists a biholomorphic map from $\sS$ to $\sS'$ that sends $B_j$ to $B'_j$ for all $j$. 

This moduli space has (real) dimension $6g+2n-6$, which can be understood simply as follows.  A hyperbolic surface $\Sigma_{g,n}(\boldsymbol{L})$ can be decomposed into $2g+n-2$ pairs of pants by cutting it along $3g+n-3$ geodesics. The hyperbolic metric on a pair of pants is uniquely determined by the lengths of the geodesics of its boundary components. This means that the moduli space $\cM_{g,n}(\boldsymbol{L})$ can be parameterized by specifying the lengths  $L_i=\{L_1, \ldots, L_n, l_1, \ldots, L_{3g+n-3}\}$ of these geodesics, along with a corresponding set of ``twist" parameters  $\vartheta_i = \{\vartheta_1, \ldots, \vartheta_n, l_1, \ldots, \vartheta_{3g+n-3}\}$ which tell us the relative angles with which these geodesics are glued together to obtain the Riemann surface. The $(L_i, \vartheta_i)$ provide a local set of coordinates on $\cM_{g,n}(\boldsymbol{L})$ known as Fenchel--Nielsen coordinates.

A given Riemann surface $\Sigma_{g,n}(\boldsymbol{L})$ can be parameterized in many different ways using this construction, since there are many different pair-of-pants decompositions of the same surface (and, moreover, each twist parameter $\vartheta_i$ is periodic with period $L_i$). The space where the Fenchel--Nielsen coordinates $(l_i, \vartheta_i)$ are allowed to be arbitrary real values (subject only to $L_i>0$) is the universal cover of the moduli space $\cM_{g,n}(\boldsymbol{L})$,  and is known as the Teichm\"{u}ller space $\cT_{g,n}(\boldsymbol{L}) $. The moduli space itself is the quotient $$\cM_{g,n}(\boldsymbol{L}) = \cT_{g,n}(\boldsymbol{L}) / \MCG(\Sigma_{g,n}(\boldsymbol{L}))$$ where $\MCG(\Sigma_{g,n}(\boldsymbol{L}))$ is the mapping class group of the Riemann surface.\footnote{The mapping class group $\MCG(\Sigma_{g,n}(\boldsymbol{L}))$ is defined to be the group of isotopy classes of orientation-preserving homeomorphisms of $\Sigma_{g,n}(\boldsymbol{L})$ to itself, fixing its boundaries setwise.  This group permutes the different pair-of-pants decompositions of the surface.}

The moduli space $\cM_{g,n}(\boldsymbol{L})$ admits a symplectic structure called the Weil--Petersson symplectic structure. It was shown by Wolpert in \cite{Wolpert:1985} that the Weil--Petersson symplectic form\footnote{This symplectic form is also valid at the level of the Teichm\"{u}ller space $\cT_{g,n}(\boldsymbol{L})$. Since this form is invariant under the action of the mapping class group $\MCG(\Sigma_{g,n}(\boldsymbol{L}))$, it will be carried over to the moduli space.} $w_{\WP}$ takes a simple form in terms of the Fenchel--Nielsen coordinates,
\begin{equation}
    w_{\WP} = \sum_{i = 1}^{3g+n-3} \diff \varrho_i \wedge \diff l_i, \qquad \varrho_i = \frac{l_i \vartheta_i}{2\pi}.
\end{equation}

One nice property of this space is that when $\boldsymbol{L}$ vanishes $\cM_{g,n}(\boldsymbol{L})$ is a K\"{a}hler manifold with respect to the Weil--Petersson symplectic structure \cite{Ahlfors:1961some,Zograf:1988hg}. This allows one to write the symplectic form in terms of the $3g + n - 3$ complex moduli $(\boldsymbol{z},\conj{\boldsymbol{z}})$\footnote{In complex $(\boldsymbol{z},\conj{\boldsymbol{z}})$-coordinates, we define the derivatives as $\partial_i \defeq \partial / \partial z_i$ and $\conj{\partial}_i \defeq \partial / \partial \conj{z}_i$.} of $\Sigma_{g,n}(\boldsymbol{L})$ as
\begin{equation}
    \omega_{\WP} = \ic \sum_{i = 1}^{3g + n - 3} \partial_i \conj{\partial}_i U \, \diff z_i \wedge \diff \conj{z}_i,
\end{equation}
where $U$ is the K\"{a}hler potential. Hence, we can write the line element corresponding to the Weil--Petersson metric as
\begin{equation}
    \diff s^2 = 2 \sum_{i = 1}^{3g + n - 3} \partial_i \conj{\partial}_i U \, \diff z_i \diff \conj{z}_i,
\end{equation}
such that $g_{i\conj{i}} = \partial_i \conj{\partial}_i U$. From here, one can write the volume $\sV_{g,n}(\boldsymbol{L})$ of the moduli space $\cM_{g,n}(\boldsymbol{L})$ as
\begin{equation}\label{eq:volgn}
    \sV_{g,n}(\boldsymbol{L}) = \int_{\cM_{g,n}(\boldsymbol{L})} \frac{\omega_{\WP}^{3g+n-3}}{(3g+n-3)!}.
\end{equation}
The existence of this K\"{a}hler structure on hyperbolic surfaces is also proven for the sphere with $n$ conical singularities in \cite{Takhtajan:2001uj}. On the other hand, in the case of bordered surfaces, the moduli space does not have a K\"{a}hler structure. Nevertheless, the prescription to compute the metric on the moduli space we will use does not depend strongly on the existence of a K\"{a}hler structure. Hence, we can carry out our procedure using Liouville theory even for surfaces with geodesic boundary components of non-vanishing length. Indeed, we will later verify that the Weil--Petersson volumes obtained using this prescription match the volume polynomials in all cases we study.

\subsection{Algebraic surfaces}\label{s:algcurves}
Certain families of Riemann surfaces (without boundary), which we will denote $\rY_{m,N}({\boldsymbol{u}},{\boldsymbol{v}})$ can be represented by an algebraic curve of the form,
\begin{equation}
    y^m = \prod_{k=1}^N \frac{z-u_k}{z-v_k},
\end{equation}
which defines an $m$-sheeted cover of the Riemann sphere $\bbC^*$ of genus $g=(N-1)(m-1)$. In the above equation, $z \in \bbC^*$ is a coordinate on the Riemann sphere, and $\{u_k,v_k\}$ are a set of $2N$ branch points which parametrize the moduli of $\rY_{m,N}$. The $m$ choices of roots $y$ correspond to the $m$ sheets in the cover, and the ``replica'' symmetry $y\to e^{2\pi \ic/m}y$ means that all surfaces written in this form have an additional $\bbZ_m$ symmetry. These surfaces have been referred to as R\'{e}nyi surfaces in the literature due to their appearance in the computation of entanglement R\'{e}nyi entropies and have been studied in, e.g., \cite{Calabrese:2009ez,Calabrese:2010he}.

As an example, we consider the family of $\bbZ_m$-symmetric genus $m-1$ surfaces $\rY_{m,2}$ defined by taking $N=2$. There exists a unique $\SL(2,\bbC)$ transformation which maps the four branch points in the following way:
$$
(u_1,v_1,u_2,v_2)\mapsto (0,x,1,\infty),
$$
where we have defined the complex cross-ratio $x$ as 
\begin{equation}
    x \defeq \frac{(u_1-v_1)(u_2-v_2)}{(u_1-u_2)(v_1-v_2)}.
\end{equation}
This family of surfaces has a one-complex-dimensional moduli space that can be parametrized by a single coordinate: the cross-ratio $x$. After performing this mapping, the defining equation for this family of $\bbZ_m$-replica surfaces becomes
\begin{equation}
    y^m=\frac{z(z-1)}{z-x}.
\end{equation}

When $m=2$, the surface $\rY_{2,2}$ defines a torus, and the complex cross-ratio $x$ is related to the usual complex structure $\tau$ of the torus by
\begin{equation}\label{eq:torusparameter}
    \tau = \ic \frac{{}_2F_1\qty(\frac{1}{2}, \frac{1}{2}; 1; 1-x)}{{}_2F_1\qty(\frac{1}{2}, \frac{1}{2}; 1; x)}, \qquad x = \qty(\frac{\Theta_2(\tau)}{\Theta_3(\tau)})^4,
\end{equation}
where $\Theta_j(\tau)$ is the $j$th elliptic theta function.

We will be interested in the case of $m=3$, which defines a family of $\bbZ_3$-symmetric genus two surfaces $\rY_{3,2}$. The moduli space of such surfaces is a one-complex-dimensional slice of the full moduli space of genus two surfaces. The period matrix\footnote{The parameter $\tilde{\tau}$ should not be confused with the parameter $\tau$ appearing in the exponent of the nome $q = \ep^{\ic \pi \tau}$, which is utilized for computational purposes. Each of these parameters can be regarded as the complex coordinate on the one-complex-dimensional moduli space.} corresponding to this family of surfaces is
\begin{equation}\label{eq:y32periodmatrix}
    \Omega = \frac{1}{\sqrt{3}} \begin{pmatrix}
        2 & -1 \\
        -1 & 2
    \end{pmatrix} \tilde{\tau}, \qquad \tilde{\tau} = \ic \frac{{}_2F_1\qty(\frac{1}{3}, \frac{2}{3}; 1; 1-x)}{{}_2F_1\qty(\frac{1}{3}, \frac{2}{3}; 1; x)}.
\end{equation}

\subsection{Volume polynomials}\label{s:volumes}
We would like to understand the Weil--Petersson volumes $\sV_{g,n}(\boldsymbol{L})$ of the moduli spaces $\cM_{g,n}(\boldsymbol{L})$ of genus $g$ Riemann surfaces with $n$ geodesic boundaries defined in equation~(\ref{eq:volgn}). It is a beautiful result due to Mirzakhani in \cite{Mirzakhani:2007} that $\sV_{g,n}(\boldsymbol{L})$ has the form of a polynomial in the lengths $\boldsymbol{L}$ whose coefficients are rational multiples of powers of $\pi$ related to the intersection numbers on moduli space. The general results of Mirzakhani match the results of prior authors in certain limits. For example, the volumes $\sV_{0,n}(0)$ of the genus 0 moduli spaces of $n$ marked points $\cM_{0,n}$ were computed by Zograf in \cite{Zograf:1993}, whereas the results for volumes of moduli spaces of general genus $g$ curves with $n$ marked points $\sV_{g,n}(0)$ (i.e., the constant terms of $\sV_{g,n}(\boldsymbol{L})$) have been computed by Wolpert in \cite{Wolpert:1983}.

The main example relevant to the surfaces we will consider in this paper is the case of $\sV_{0,4}(\boldsymbol{L})$, the volume of the moduli space of genus 0 surfaces with four geodesic boundaries. The result is
\begin{equation}\label{eq:4bpolynomial}
    \sV_{0,4}(\boldsymbol{L}) = 2\pi^2 + \frac{1}{2}\qty(L_1^2 + L_2^2 + L_3^2 + L_4^2).
\end{equation}
When we take all the lengths to zero, we find the volume of the moduli space of the four-punctured sphere, $\sV_{0,4}= 2\pi^2$.

Though the work of Mirzakhani applies to Riemann surfaces of general genus $g$ with either geodesic boundaries or punctures, she suggests in \cite{Mirzakhani:2007} that it may be easy to generalize her results to the case of cone points of deficit angles $\theta_i \in [0, \pi)$ via analytic continuation of the geodesic boundary lengths to complex values $L_i=-\ic\theta_i$. This is easily seen to be true for the case of $\sV_{0,\conj{4}}(\theta_1,\theta_2,\theta_3,\theta_4)$,\footnote{Here we temporarily denote the volume of the moduli space of a genus $g$ surface with $n$ conical defects as $\sV_{g,\conj{n}}$ to emphasize the fact that we are considering surfaces with cone point singularities rather than geodesic boundaries and that the volumes of these two types of moduli space may not be related a priori.} the volume of the moduli space of genus 0 surfaces with four cone points of deficit angles $\theta_1,...,\theta_4$, $\theta_i \in [0, \pi)$, which was analyzed in \cite{Nakanishi:2001ar} and found to be
\begin{equation}\label{eq:4cpolynomial}
    \sV_{0,\conj{4}}(\boldsymbol{\theta})=\sV_{0,4}(-\ic\boldsymbol{\theta})=2\pi^2 - \frac{1}{2} \qty(\theta_1^2 + \theta_2^2 + \theta_3^2 + \theta_4^2).
\end{equation}
This is shown to be true in general in \cite{Tan:2006}; thus, one can always deduce the volumes of moduli spaces hyperbolic Riemann surfaces including conical singularities of deficit angles $\theta_i \in [0, \pi)$ by analytically continuing the lengths appearing in Mirzakhani's formula to $L_i = -\ic \theta_i$. We will use this fact when we study volumes of moduli spaces of genus zero Riemann surfaces with four conical deficits in \S\ref{s:examples}. When the deficit angles are not in this range, the structure of the analytic continuation is somewhat more subtle, but in some cases, moduli space volumes can still be considered (see e.g. \cite{Eberhardt:2023rzz,Artemev:2023bqj} for a recent discussion).

\section{Liouville theory}\label{s:liouville}
In this section, we briefly review some aspects of quantum Liouville theory, its observables, and their connections to the geometry of hyperbolic surfaces. The main purpose of this section is to introduce the notation that we use later in the paper. For more complete reviews of Liouville theory see \cite{Seiberg:1990eb,Teschner:2001rv,Nakayama:2004vk}.

\subsection{Liouville conformal field theory}
Liouville theory is a two-dimensional quantum field theory with the following classical action:
\begin{equation}\label{eq:liouville:action}
    S^{\Sigma}_{\rL}[\hat{g},\phi] = \frac{1}{4 \pi} \int_{\Sigma} \diff^2\xi \, \sqrt{\hat{g}} \qty(\hat{g}^{\rho \sigma} \partial_\rho \phi \partial_\sigma \phi + 4 \pi \mu \ep^{2b\phi}),
\end{equation}
on a Riemann surface $\Sigma$ with coordinates\footnote{The integral measure in $\xi$-coordinates is defined as $\diff^2\xi \defeq \diff \xi^1 \wedge \diff \xi^2$.} $\xi^\rho$, where $\mu$ is the Liouville cosmological constant. The metric $\hat{g}$ is a background metric, which we will set to flat with the following line element:
\begin{equation}
    \diff \hat{s}^2 = \diff z \diff \conj{z},
\end{equation}
in complex coordinates $(z,\conj{z})$.

At the quantum level, Liouville theory is a conformal field theory (CFT) that fits into the BPZ framework \cite{Belavin:1984vu} with a continuous spectrum whose space of states decomposes as
\begin{equation}
    \cH = \int_{\bbS_Q} \diff \alpha \, \cV_{\alpha,c} \otimes \cV_{\alpha,c}, \quad \bbS_Q = \frac{Q}{2} + \ic \bbR^+,
\end{equation}
where $\cV_{\alpha,c}$ is an irreducible representation of the Virasoro algebra with central charge $c = 1 + 6Q^2$ and primary dimension $\Delta_\alpha = \alpha(Q-\alpha)$, with $Q = b+b^{-1}$. 

In Liouville CFT, we are interested in the correlation functions of primary vertex operators $V_\alpha(z,\conj{z})$ with dimension $\Delta_\alpha$, where
\begin{equation}
    \alpha = \frac{Q}{2} + \ic P, \quad P \in \bbR^+.
\end{equation}
In this notation, we refer to $P$ as the Liouville momentum. Just as in any other CFT, the two-point functions of primary operators in Liouville theory are fixed by conformal symmetry. The three-point functions take the form,
\begin{equation}
    \expval{V_{\alpha_3}(z_3, \conj{z}_3) V_{\alpha_2}(z_2, \conj{z}_2) V_{\alpha_1}(z_1, \conj{z}_1)} = \abs{z_{12}}^{2\Delta_{12,3}} \abs{z_{23}}^{2\Delta_{23,1}} \abs{z_{31}}^{2\Delta_{31,2}} C(\alpha_3, \alpha_2, \alpha_1),
\end{equation}
where $z_{ij} = z_i - z_j$ and $\Delta_{ij,k} = \Delta_k - \Delta_i - \Delta_j$ for $i,j,k \in \{1,2,3\}$ and $i \neq j \neq k \neq i$. A nontrivial fact about Liouville CFT is that its structure coefficients $C(\alpha_1,\alpha_2,\alpha_3)$ have been determined by imposing the constraints of conformal invariance and crossing symmetry and are given by the DOZZ formula first suggested in \cite{Dorn:1994xn,Zamolodchikov:1995aa} and later verified in \cite{Ponsot:1999uf}.

A generic $n$-point Liouville correlator on a Riemann surface $\Sigma_g$ of genus $g$ can be expressed as an integral over $\varsigma = 3g + n - 3$ exchange momenta of a holomorphically factorized expression that involves the structure coefficients and conformal blocks,
\begin{equation}\label{eq:liouville:npointfn}
    \expval{V_{\alpha_n}(z_n, \conj{z}_n) \cdots V_{\alpha_1}(z_1, \conj{z}_1)}_{\Sigma_g} = \int_{\bbS_Q^\varsigma} \diff \boldsymbol{\beta} \, \hat{C}^{\Sigma_g}(\boldsymbol{\alpha},\boldsymbol{\beta}) \cF^{\Sigma_g}(\boldsymbol{\alpha},\boldsymbol{\beta};\boldsymbol{z}) \cF^{\Sigma_g}(\boldsymbol{\alpha},\boldsymbol{\beta};\conj{\boldsymbol{z}}),
\end{equation}
where $\boldsymbol{\beta} = (\beta_1,\dots,\beta_\varsigma)$ parametrizes the exchange momenta, $\boldsymbol{z} = (z_1,\dots,z_\varsigma)$ denotes the moduli, and $\cF^{\Sigma_g}(\boldsymbol{\alpha},\boldsymbol{\beta};\boldsymbol{z})$ and $\cF^{\Sigma_g}(\boldsymbol{\alpha},\boldsymbol{\beta};\conj{\boldsymbol{z}})$ are the holomorphic and anti-holomorphic $n$-point conformal blocks on $\Sigma_g$ respectively. In this expression, $\hat{C}^{\Sigma_g}(\boldsymbol{\alpha},\boldsymbol{\beta})$ is the product of all the contributing three-point coefficients,\footnote{Some authors may choose to incorporate the three-point coefficients into the definition of the conformal blocks.}
\begin{equation}\label{eq:liouville:3pointcont}
    \hat{C}^{\Sigma_g}(\boldsymbol{\alpha},\boldsymbol{\beta}) = \prod_{k = 1}^{2g+n-2} C({\lambda_k}_1,{\lambda_k}_2,{\lambda_k}_3), \quad {\lambda_k}_i \in \{\alpha_1, \dots, \alpha_n, \beta_1, \dots, \beta_\varsigma, Q-\beta_1, \dots, Q-\beta_\varsigma\}.
\end{equation}

Implicit in formula~(\ref{eq:liouville:npointfn}) is a choice of conformal block decomposition of Liouville correlators on the Riemann surface $\Sigma_g$, corresponding to a pants decomposition of $\Sigma_g$ with a complete set of primary states inserted at each cut. This choice also specifies the three momenta appearing in each structure coefficient of the product in equation~(\ref{eq:liouville:3pointcont}). In this picture, each three-point structure coefficient corresponds to the contribution of the corresponding pair of pants, and the integral over the exchange momenta represents the sum over the set of primary states at each cut. The conformal blocks then include the contributions of the descendants for each of these primary states.

As an example, the conformal block decomposition for the four-point function on the sphere is
\begin{equation}
    \expval{V_{\alpha_4}(\infty) V_{\alpha_3}(1) V_{\alpha_2}(x,\conj{x}) V_{\alpha_1}(0)} = \int_{\bbS_Q} \diff \beta \, C(\alpha_4,\alpha_3,\beta) C(Q-\beta,\alpha_2,\alpha_1) \cF_{34}^{21}(\beta; x) \cF_{34}^{21}(\beta; \conj{x}),
\end{equation}
where we have used a conformal transformation to introduce the cross-ratio, $x$, defined by
\begin{equation}
x:= \frac{(z_1-z_2)(z_3-z_4)}{(z_1-z_4)(z_3-z_2)},
\end{equation}
and $\cF_{34}^{21}(\beta; x) \defeq \cF^{\Sigma_{0,4}}(\boldsymbol{\alpha}, \beta; x)$ denotes the four-point Virasoro blocks on the sphere. Here we have introduced the cross-ratio $x$, 
These conformal blocks are not known in closed form for arbitrary (external and exchange) momenta. However, they can be obtained as an expansion in the cross-ratio $x$ using Zamolodchikov's recursion relations \cite{Zamolodchikov:1987}. Notice that in this formula, the second three-point structure includes the insertion of a reflected vertex operator $V_{Q-\beta} = R(\beta) V_\beta$, i.e.,
\begin{equation}
    C(Q-\beta,\alpha_2,\alpha_1) = R(\beta) C(\beta,\alpha_2,\alpha_1),
\end{equation}
where $R(\beta)$ is called the reflection coefficient. 

Another example is the genus two partition function,
\begin{equation}
    \cZ^{\Sigma_{2,0}}(\boldsymbol{q},\conj{\boldsymbol{q}}) = \int_{\bbS_Q^3} \diff \boldsymbol{\beta} \, C(\beta_1,\beta_2,\beta_3) C(Q-\beta_1,Q-\beta_2,Q-\beta_3) \cF^{\Sigma_{2,0}}(\boldsymbol{\beta}; \boldsymbol{q}) \cF^{\Sigma_{2,0}} (\boldsymbol{\beta};\conj{\boldsymbol{q}}),
\end{equation}
where $\boldsymbol{q} = (q_1,q_2,q_3)$ and $\conj{\boldsymbol{q}} = (\conj{q}_1,\conj{q}_2,\conj{q}_3)$ are the coordinates that parametrize the moduli space of genus two Riemann surfaces and $\cF(\boldsymbol{\beta}; \boldsymbol{q})$ are the genus two conformal blocks \cite{Cho:2017oxl,Cho:2017fzo} with exchange momenta $\boldsymbol{\beta} = (\beta_1, \beta_2, \beta_3)$.

\subsection{Semiclassical limit of Liouville theory path integral}\label{s:semiclass}
In the semiclassical limit with $b \to 0$, Liouville theory describes the classical geometry of hyperbolic Riemann surfaces. To see this, we can introduce a new field $\varphi \defeq 2b\phi$. The action for the field $\varphi$ in flat background metric with complex $(z,\conj{z})$-coordinates\footnote{The integral measure in complex $(z,\conj{z})$-coordinates is defined as $\diff^2z \defeq \frac{1}{2\ic} \diff z \wedge \diff \conj{z}$.} on a surface $\Sigma$ is
\begin{equation}
    S^{\Sigma}_{\rL}[\varphi] = b^{-2} S^{\Sigma}_{\rmc}[\varphi] = \frac{1}{4\pi b^2} \int_\Sigma \diff^2z \, \qty(\partial \varphi \conj{\partial} \varphi + 4\pi \lambda \ep^{\varphi}),
\end{equation}
where $\lambda \defeq \mu b^2$ is fixed as $b \to 0$. The classical equation of motion for $\varphi$,
\begin{equation}
    \partial \conj{\partial} \varphi = 2\pi \lambda \ep^\varphi,
\end{equation}
is the Liouville equation. The solution to this equation furnishes the conformal factor of the physical metric $g_{\mu\nu} = \ep^\varphi \hat{g}_{\mu\nu}$ on the Riemann surface $\Sigma$, with the following line element:
\begin{equation}\label{eq:liouville:metric}
    \diff s^2 = \ep^{\varphi} \diff z \diff \conj{z}.
\end{equation}
Note that the Gaussian curvature for this metric is given by
\begin{equation}
    K = - 2 \ep^{-\varphi} \partial \conj{\partial} \varphi,
\end{equation}
so the solution to the Liouville equation yields a metric with constant negative curvature $K = -4 \pi \lambda$. On the Riemann sphere, in order to ensure that the physical metric is smooth everywhere, we need to impose the following asymptotic behavior for the field $\varphi$ at infinity:
\begin{equation}\label{eq:liouville:bc_inf}
    \varphi(z,\conj{z}) = -2 \log z\conj{z} + \cO\qty(1) \quad \text{as} \quad \abs{z} \to \infty.
\end{equation}

The primary vertex operators in Liouville theory are of the form $V_\alpha = \ep^{2\alpha \phi}$. The correlation function of such operators can be written in the path integral language as
\begin{equation}
    \expval{V_{\alpha_n}(z_n,\conj{z}_n) \dots V_{\alpha_1}(z_z,\conj{z}_1)} = \int \Diff \varphi \, \ep^{-b^{-2} S_\rmc[\varphi]} \prod_{i=1}^n \exp(\frac{\alpha_i \varphi(z_i,\conj{z}_i)}{b}).
\end{equation}
Since we are looking for the saddle-point approximation for this path integral for $b \to 0$, we must understand how each $\alpha_i$ scales with $b$. For an operator to have a nontrivial effect on the saddle point, we need $\alpha$ to scale with $b^{-1}$. Thus, we take $\alpha = b^{-1} \eta$ and keep $\eta$ fixed as $b \to 0$ for an operator that affects the saddle point. Such operators are called heavy primaries. Asymptotically, heavy operators have dimension $\Delta = b^{-2} \eta (1-\eta)$ in the limit $b \to 0$.

Semiclassically, the insertion of heavy operators adds delta function terms to the action, resulting in a modified equation of motion:
\begin{equation}\label{eq:liouville:insertion}
    \partial \conj{\partial} \varphi = 2 \pi \lambda \ep^{\varphi} - 2\pi \sum_i \eta_i \delta^{(2)}(z - z_i),
\end{equation}
where $\delta^{(2)}(z - z_i)$ is the Dirac delta function in two dimensions. This implies that in the neighborhood of a heavy operator insertion, we have
\begin{equation}
    \varphi(z,\conj{z}) = -2\eta_i \log{\abs{z - z_i}^2} \quad \text{as} \quad \abs{z - z_i} \to 0.
\end{equation}

In principle, one can find the $\varphi^{\rms}$ that solves equation~(\ref{eq:liouville:insertion}) globally on a Riemann surface $\Sigma$, parameterized by a number of moduli including the period matrix and insertion points, and compute the critical Liouville action as $S^{\Sigma}_{\rmc}[\varphi^{\rms}]$. The interpretation of Liouville field $\varphi^{\rms}$ as the conformal factor of the metric on a surface $\Sigma$ in equation~(\ref{eq:liouville:metric}) allows us to think of the critical Liouville action as only a function of the moduli $(\boldsymbol{q}, \boldsymbol{\conj{q}})$ of $\Sigma$,
\begin{equation}
    S^{\Sigma}_{\rmc}(\boldsymbol{q}, \conj{\boldsymbol{q}}) \defeq S^{\Sigma}_{\rmc}[\varphi^{\rms}].
\end{equation}
Remarkably, this Liouville action has a simple interpretation in terms of the Weil--Petersson symplectic form: it is proportional to the K\"{a}hler potential on moduli space, $U = - 4 \pi S_{\rmc}$ \cite{Zograf:1988hg,Takhtajan:2001uj}.

\subsection{Semiclassical limit of the conformal block expansion}\label{s:blocks}
In the semiclassical limit, the DOZZ formula for the structure coefficients $C(\alpha_1,\alpha_2,\alpha_3)$ can be written in an exponential form \cite{Zamolodchikov:1995aa,Harlow:2011ny},
\begin{equation}\label{eq:liouville:dozzexp}
    \lim_{b \to 0} C(\alpha_1,\alpha_2,\alpha_3) = \exp(- b^{-2} S_3(p_1,p_2,p_3)),
\end{equation}
where $S_3(p_1,p_2,p_3)$ is equal to the critical Liouville action on a sphere with three heavy operator insertions (see \S\ref{a:semiclassics}). Similarly, the reflection coefficients exponentiate in this limit,
\begin{equation}\label{eq:liouville:refexp}
    \lim_{b \to 0} R(\alpha) = \exp(-b^{-2} r(p)).
\end{equation}
In these expressions, we have used the following parametrization for the Liouville momenta:
\begin{equation}
    \alpha_i = \frac{Q}{2} + \ic P_i, \quad \eta_i = \frac{1}{2} + \ic p_i,
\end{equation}
such that in the semiclassical limit $\alpha_i = b^{-1} \eta_i$ and $P_i = b^{-1} p_i$.

Virasoro blocks of heavy operators also take an exponential form in the limit $b \to 0$. This was conjectured \cite{Zamolodchikov:1987} and proven \cite{Besken:2019jyw} for the sphere four-point blocks $\cF_{34}^{21}(\alpha; x)$. In this limit,
\begin{equation}
    \lim_{b \to 0} \cF_{34}^{21}(\alpha; x) = \exp(b^{-2} f_{34}^{21}(p; x)).
\end{equation}
General Virasoro blocks for an arbitrary number of heavy operator insertions are also believed to behave similarly in the semiclassical limit, e.g., for genus two blocks:
\begin{equation}
    \cF^{\Sigma_{2,0}}(\boldsymbol{\alpha}; \boldsymbol{q}) = \exp(b^{-2} f^{\Sigma_{2,0}}(\boldsymbol{p}; \boldsymbol{q})).
\end{equation}
This exponentiation property allows us to make a saddle-point approximation to the conformal block expansion of an observable in Liouville theory on higher genus surfaces.

In the case of a sphere four-point correlator, we can write
\begin{equation}\label{eq:sphere-4pt}
    \lim_{b \to 0} \expval{V_{\alpha_4}(\infty) V_{\alpha_3}(1) V_{\alpha_2}(x,\conj{x}) V_{\alpha_1}(0)} \approx \int_\bbR \diff p \, \exp(- b^{-2} S_{34}^{21}(p; x,\conj{x})),
\end{equation}
where
\begin{equation}
    S_{34}^{21}(p; x,\conj{x}) = S_3(p_4,p_3,p) + r(p) + S_3(p,p_2,p_1) - f_{34}^{21}(p; x) - f_{34}^{21}(p; \conj{x}).
\end{equation}
In the semiclassical limit, we can find the saddle-point momentum $p^{\rms}$ from the saddle-point equation
\begin{equation}
    \eval{\frac{\partial S_{34}^{21}(p;x,\conj{x})}{\partial p}}_{p = p^{\rms}} = 0,
\end{equation}
and interpret the value of the exponent as the critical Liouville action,
\begin{equation}
    S^{\Sigma_{0,4}}_{\rmc}(x,\conj{x}) = S_{34}^{21}(p^{\rms};x,\conj{x}).
\end{equation}

Similarly, for the genus two partition function, we can write the conformal block expansion as
\begin{equation}
    \lim_{b \to 0} \cZ^{\Sigma_{2,0}}(\boldsymbol{q},\conj{\boldsymbol{q}}) \approx \int_{\bbR^3} \diff^3p \, \exp(-b^{-2} S^{\Sigma_{2,0}}(p_1,p_2,p_3;\boldsymbol{q},\conj{\boldsymbol{q}})),
\end{equation}
where
\begin{equation}
    \begin{split}
        S^{\Sigma_{2,0}}(p_1,p_2,p_3;\boldsymbol{q},\conj{\boldsymbol{q}}) &= 2 S_3(p_1,p_2,p_3) + r(p_1) + r(p_2) + r(p_3) \\
        &\quad- f^{\Sigma_{2,0}}(p_1,p_2,p_3;\boldsymbol{q}) - f^{\Sigma_{2,0}}(p_1,p_2,p_3; \conj{\boldsymbol{q}}).
    \end{split}
\end{equation}
In this case, we have a set of saddle-point equations,
\begin{equation}
    \eval{\frac{\partial S^{\Sigma_{2,0}}(p_1,p_2,p_3;\boldsymbol{q},\boldsymbol{q})}{\partial p_i}}_{p_j = p^{\rms}_j} = 0, \quad i, j \in \{1,2,3\},
\end{equation}
that we should solve for $p^{\rms}_j$, in order to find the critical Liouville action,
\begin{equation}
    S^{\Sigma_{2,0}}_{\rmc}(\boldsymbol{q},\conj{\boldsymbol{q}}) = S^{\Sigma_{2,0}}\qty(p^{\rms}_1,p^{\rms}_2,p^{\rms}_3;\boldsymbol{q},\conj{\boldsymbol{q}}).
\end{equation}
The value of the saddle-point momenta $p_i^{\rms}$ are related to the length $l_i$ of the geodesic cuts in the pair of pants decomposition of our hyperbolic surface via $l_i = 4\pi p_i^{\rms}$. Hence, finding their values is equivalent to finding the constant negative curvature metric on the surface.

\section{The Weil--Petersson metric}\label{s:metric}
As explained previously, the critical Liouville action, which furnishes---conjecturally, in some cases (see \S\ref{s:semiclass})---a K\"{a}hler potential on the moduli space, can be calculated by solving the saddle-point equation for the conformal block expansion of an observable in Liouville theory. However, obtaining closed-form solutions for such saddle-point equations is generally impossible for two reasons. Firstly, conformal blocks usually can only be expressed as perturbative expansions in moduli $\boldsymbol{q}$ and do not take a closed-form expression except for particular cases. Secondly, the saddle-point equations must be solved for the $\varsigma = 3g + n - 3$ exchange momenta $\boldsymbol{p} = (p_1, \dots, p_\varsigma)$ that appear in the arguments of multiple Gamma functions resulting from the DOZZ formula. Analytically solving such equations is generally impossible even to lowest orders in the $\boldsymbol{q}$-expansions of the conformal block.\footnote{In \cite{Harrison:2022frl}, the case of the one-complex-dimensional moduli space $\cM_{0,4}$ was studied, where it was shown that at a fixed order in the $q$-expansion of the conformal block, one could identify a small parameter $\varepsilon(q,\conj{q})$ which allows for series expansion solutions to the saddle-point equation, i.e.,
\begin{equation}
    p^{\rms} = \sum_{n=1} p^{\rms}_{n} \varepsilon(q,\conj{q})^n,
\end{equation}
where $p = p^{\rms}$ is the solution to the saddle-point equations. However, trying to adapt such methods to other moduli spaces results in non-convergent expansions in many cases, including those we study in this paper. Thus, we use a different method here.}

To compute the critical Liouville action and metric on the moduli space, we will utilize numerical methods to solve the saddle-point equation. However, we will not rely on numerical methods for differentiation. We will provide a detailed description of this approach in this section.

\subsection{Computing the metric on moduli space}
The Weil--Petersson metric on the moduli space $\cM_{g,n}(\boldsymbol L)$ parametrized by $(\boldsymbol{q}, \conj{\boldsymbol{q}})$ is obtained from the Liouville critical action $S_{\rmc}(\boldsymbol{q},\conj{\boldsymbol{q}})$ as
\begin{equation}
    g_{i\conj{i}} = - 4\pi \frac{\partial^2 S_{\rmc}(\boldsymbol{q}, \conj{\boldsymbol{q}})}{\partial q_i \partial \conj{q}_i}.
\end{equation}
In the saddle-point approximation, the critical Liouville action can be written in full generality as
\begin{equation}
    S_{\rmc}(\boldsymbol{q}, \conj{\boldsymbol{q}}) = S(\boldsymbol{p}^{\rms}; \boldsymbol{q}, \conj{\boldsymbol{q}}) = \hat{S}_3(\boldsymbol{p}^{\rms}) - \hat{f}(\boldsymbol{p}^{\rms}; \boldsymbol{q}) - \hat{f}(\boldsymbol{p}^{\rms}; \conj{\boldsymbol{q}}),
\end{equation}
where we reiterate that $\boldsymbol{p}^{\rms}$ is a $(3g+n-3)$-dimensional vector of saddle-point exchange momenta depending on the choice of pants decomposition of $\Sigma_{g,n}(\boldsymbol L)$.
In the above, $\hat{S}_3(\boldsymbol{p}^{\rms})$ denotes the total contribution of the three-point functions, including the reflection coefficients, whereas $\hat{f}(\boldsymbol{p}^{\rms}; \boldsymbol{q})$ and $\hat{f}(\boldsymbol{p}^{\rms}; \conj{\boldsymbol{q}})$ denote the total contribution of holomorphic and anti-holomorphic conformal blocks respectively. 

The saddle-point momentum is then computed by solving the following saddle-point equation,
\begin{equation}\label{eq:pro:saddle}
    \eval{\frac{\partial S(\boldsymbol{p}; \boldsymbol{q}, \conj{\boldsymbol{q}})}{\partial p_j}}_{\boldsymbol{p} = \boldsymbol{p}^{\rms}} = \qty[\frac{\partial \hat{S}_3(\boldsymbol{p})}{\partial p_j} - \frac{\partial \hat{f}(\boldsymbol{p}; \boldsymbol{q})}{\partial p_j} - \frac{\partial \hat{f}(\boldsymbol{p}; \conj{\boldsymbol{q}})}{\partial p_j}]_{\boldsymbol{p} = \boldsymbol{p}^{\rms}} = 0.
\end{equation}
In this setup, we can write
\begin{align}
    \frac{\partial^2 \hat{f}(\boldsymbol{p}^{\rms}; \boldsymbol{q})}{\partial q_i \partial \conj{q}_i} &= \frac{\partial}{\partial q_i} \qty[\frac{\partial p^{\rms}_j}{\partial \conj{q}_i} \frac{\partial \hat{f}(\boldsymbol{p}; \boldsymbol{q})}{\partial p_j}]_{\boldsymbol{p}=\boldsymbol{p}^{\rms}}, \\
    \frac{\partial^2 \hat{f}(\boldsymbol{p}^{\rms}; \conj{\boldsymbol{q}})}{\partial q_i \partial \conj{q}_i} &= \frac{\partial}{\partial \conj{q}_i} \qty[\frac{\partial p^{\rms}_j}{\partial q_i} \frac{\partial \hat{f}(\boldsymbol{p}; \conj{\boldsymbol{q}})}{\partial p_j}]_{\boldsymbol{p}=\boldsymbol{p}^{\rms}},
\end{align}
and
\begin{equation}
    \frac{\partial^2 \hat{S}_3(\boldsymbol{p}^{\rms})}{\partial q_i \partial \conj{q}_i} = \frac{\partial}{\partial \conj{q}_i} \qty[\frac{\partial p^{\rms}_j}{\partial q_i} \frac{\partial \hat{S}_3(\boldsymbol{p})}{\partial p_j}]_{\boldsymbol{p}=\boldsymbol{p}^{\rms}}.
\end{equation}
Combining these together and using equation~(\ref{eq:pro:saddle}), we find
\begin{equation}
    \frac{\partial^2 S_{\rmc}(\boldsymbol{q}, \conj{\boldsymbol{q}})}{\partial q_i \partial \conj{q}_i} = \qty[\frac{\partial p^{\rms}_j}{\partial \conj{q}_i} \frac{\partial^2 \hat{f}(\boldsymbol{p}; \boldsymbol{q})}{\partial q_i \partial p_j}]_{\boldsymbol{p}=\boldsymbol{p}^{\rms}}.
\end{equation}
On the other hand, by taking a derivative of equation~(\ref{eq:pro:saddle}) with respect to $\conj{q}_i$, we find
\begin{equation}
    \frac{\partial p^{\rms}_j}{\partial \conj{q}_i} = \qty[\qty(\frac{\partial^2 S(\boldsymbol{p}; \boldsymbol{q}, \conj{\boldsymbol{q}})}{\partial p_k \partial p_j})^{-1} \frac{\partial^2 \hat{f}(\boldsymbol{p};\conj{\boldsymbol{q}})}{\partial \conj{q}_i \partial p_k}]_{\boldsymbol{p} = \boldsymbol{p}^{\rms}},
\end{equation}
which results in
\begin{equation}\label{eq:pro:metricfromps}
    g_{i \conj{i}} = 4\pi \qty[\qty(\frac{\partial^2 S(\boldsymbol{p}; \boldsymbol{q}, \conj{\boldsymbol{q}})}{\partial p_k \partial p_j})^{-1} \frac{\partial^2 \hat{f}(\boldsymbol{p}; \boldsymbol{q})}{\partial q_i \partial p_j} \frac{\partial^2 \hat{f}(\boldsymbol{p};\conj{\boldsymbol{q}})}{\partial \conj{q}_i \partial p_k}]_{\boldsymbol{p} = \boldsymbol{p}^{\rms}}.
\end{equation}
This formula implies that in order to compute the Weil--Petersson metric, even in the cases where it is not admissible to perform a Taylor expansion in momentum variables, the only part of the computation that we must do numerically is finding the solution to the saddle-point equation~(\ref{eq:pro:saddle}).

\subsection{Computing the volume of moduli space}
In the following section, we will be concerned with integrating the Weil--Petersson metric and finding the corresponding volume of moduli space. We will specialize to the case of the sphere with four singularities and for simplicity, we will focus on the case where all four singularities are identical, this is the moduli space $\cM_{0,4}(L)$.

Let $\{z_1, z_2, z_3, z_4\} \subset  \bbC^*$ denote the locations of the four singularities on the Riemann sphere. There exists a unique M\"{o}bius transformation which sends these four points to $\{x, 0, 1, \infty\} \subset  \bbC^*$. The coordinate $x$ resulting from this transformation, known as the cross-ratio, is a complex coordinate on the moduli space $\cM_{0,4}(L)$. The cross-ratio is explicitly given by the formula
\begin{equation}
    x = \frac{(z_1 - z_2)(z_3 - z_4)}{(z_1 - z_4)(z_2 - z_3)}.
\end{equation}

Due to the four singularities being identical, the moduli space naturally admits the action of the group $S_4$ which acts by permuting the four points. The cross-ratio is invariant under the subgroup $ \bbZ_2 \times \bbZ_2$ which acts by transposing the four points pairwise. The remaining permutations form the anharmonic group $S_4 / ( \bbZ_2 \times  \bbZ_2) \cong S_3$. This group acts on the cross-ratio as the transformations,
\begin{equation}
    S_3: \quad x \rightarrow \qty{x, 1-x, \frac{1}{x}, \frac{1}{1-x}, \frac{x}{1-x}, \frac{x-1}{x}}.
\end{equation}

\begin{figure}
    \centering
    \begin{subfigure}{.47\textwidth}
        \centering
        \begin{tikzpicture}
            \pgfmathsetmacro\sqrtthree{sqrt(3)}
            \draw[dashed] (1,-2.6) -- (1,2.6);
            \draw[dashed] (-2.2,0) -- (4.2,0);
            \draw[dashed] (4,0) arc (0:180:2) node[anchor=north east]{$0$};
            \draw[dashed] (4,0) arc (0:-180:2);
            \draw[dashed] (2,0) node[anchor=north west]{$1$} arc (0:360:2);
            \draw (0,0) -- (1,0) -- (1,\sqrtthree);
            \draw (1,\sqrtthree) arc (120:180:2);
            \fill[gray!23, even odd rule] (0,0) -- (1,0) -- (1,\sqrtthree) arc (120:180:2);
            \draw (4.2,2.6) -- (3.8,2.6) node[anchor = south west]{$x$} -- (3.8,3);
        \end{tikzpicture}
        \caption{}\label{fig:funDomainX}
        \end{subfigure}\hspace*{\fill}
        \begin{subfigure}{.47\textwidth}
        \centering
        \begin{tikzpicture}
            \pgfmathsetmacro\sqrtthree{sqrt(3)}
            \draw (-3,0) -- (-2,0) node[anchor=north]{$-1$} -- (0,0) node[anchor=north]{$0$} -- (2,0) node[anchor=north]{$1$} -- (3,0);
            \draw[dashed] (-2,0) -- (-2,5);
            \draw[dashed] (-1,1) -- (-1,5);
            \draw[dashed] (0,0) -- (0,5);
            \draw[dashed] (1,1) -- (1,5);
            \draw[dashed] (2,0) -- (2,5);
            \draw[dashed] (0,0) arc (0:180:1);
            \draw[dashed] (0,0) arc (180:0:1);
            \draw[dashed] (0,0) arc (0:90:2);
            \draw[dashed] (0,0) arc (180:90:2);
            \draw[dashed] (2,0) arc (0:180:2);
            \draw (0,5) -- (0,2);
            \draw (0,2) arc (90:60:2);
            \draw (1,\sqrtthree) -- (1,5);
            \fill[gray!23, even odd rule] (0,5) -- (0,2) arc (90:60:2) -- (1,5);
            \draw (3,4.6) -- (2.6,4.6) node[anchor = south west]{$\tau$} -- (2.6,5);
        \end{tikzpicture}
        \caption{}\label{fig:funDomainT}
        \end{subfigure}
    \caption{The half fundamental domain used for the integration of the  Weil--Petersson metric in (a) the cross-ratio coordinate, $x$, and (b) the $\tau$ coordinate.}
\end{figure}
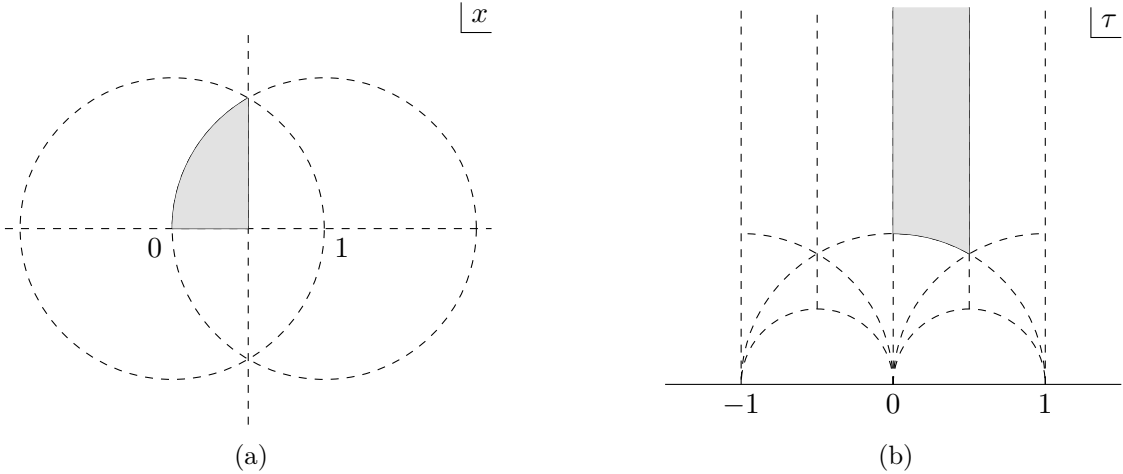

The action of the anharmonic group divides the complex plane into six fundamental domains. We can then view the moduli space $\cM_{0,4}(L)$ as a $6$-fold cover of one of these fundamental domains. The fundamental domain that we will use in this paper is
\begin{equation}
    D = \qty{x \in  \bbC \;\middle|\; \Re x < \frac{1}{2}, \abs{x-1} < 1}.
\end{equation}
This choice of fundamental domain is convenient for our calculations as we will be using approximations for the conformal blocks that are valid near $x=0$. The metric has an additional $ \bbZ_2$ symmetry which is the action of complex conjugation, explicitly $x \leftrightarrow \bar{x}$. This leads to the half fundamental domain in figure~\ref{fig:funDomainX}, which is given by
\begin{equation}
    D_{\text{half}} = \qty{x \in \bbC \;\middle|\; \Re x < \frac{1}{2}, \Im x > 0, \abs{x-1}  < 1}.
\end{equation}

For numerical calculations, we will find more convenience using the coordinate $\tau \in  \bbH$ that is related to the cross-ratio through the formula
\begin{equation}
    \tau = \ic \frac{K(1-x)}{K(x)}.
\end{equation}
where $K(x)$ is the complete elliptic integral of the first kind, defined by
\begin{equation}
    K(x) = \int_0^1 \frac{\diff t}{\sqrt{(1-t^2)(1-xt^2)}}.
\end{equation}
This $\tau$ coordinate is related to Zamolodchikov's $q$ coordinate by $q = \exp(\pi\ic\tau)$ and the utility of using $\tau$ comes from the fact that the four-point Virasoro block converges more rapidly when expressed in this $q$ coordinate.  The result is that the Zamolodchikov recursion relations will provide an excellent approximation to the conformal block including only the first few terms in the $q$-expansion. The half fundamental domain for $\tau$ is shown in figure~\ref{fig:funDomainT}.

\section{Examples}\label{s:examples}
In this section, we provide examples of one-dimensional moduli spaces to demonstrate the applications of the proposed method for calculating Weil--Petersson metrics. In the first example (\S\ref{s:M04}), we study the geometry of a sphere with four singularities (boundaries, punctures, or conical defects) and compute the geodesic length of the pants cut. By numerically computing the critical Liouville action to determine the K\"{a}hler potential on the moduli space, we obtain the Weil--Petersson metric and compute the corresponding moduli space volume. We compare the volumes with the infamous volume polynomials, reviewed in \S\ref{s:volumes}, demonstrating the efficacy of our method.

In the second example (\S\ref{s:genus2Z3}), we consider the $\bbZ_3$-symmetric genus two surfaces described in \S\ref{s:algcurves} and compute the volume of their moduli space in addition to the geodesic lengths of the three pants cuts. We find that the volume of the moduli space of such surfaces is three times that of a sphere with four conical singularities, each characterized by a deficit angle of $\frac{\pi}{3}$, an expectation we explain below. We comment on our results in the end.

\subsection{Sphere with four singularities}\label{s:M04}
We start with the example of a sphere with four singularities. In the language of Liouville theory, this setup corresponds to the four-point function $\expval{V_{\alpha_4}(\infty) V_{\alpha_3}(1) V_{\alpha_2}(x,\conj{x}) V_{\alpha_1}(0)}$  of heavy operators $V_{\alpha_i}$ on the sphere. Following the notation in equation~(\ref{eq:sphere-4pt}), in the semiclassical limit,
\begin{equation}\label{eq:ex:dimconv}
    \lim_{b \to 0} b \alpha_i = \frac{1}{2} + \ic p_i,
\end{equation}
where $p_i$ is the Liouville momentum of the operator $V_{\alpha_i}$. In this setup, we can consider three different scenarios (parallel to the three types of singularities mentioned in \S\ref{s:hypergeom}):
\begin{itemize}
    \item The case with $p_i \in \bbR^+$ correspond to the insertion of primary vertex operators. These operators have the property that $p_i^2 > 0$ and correspond to singularities that are geodesic boundaries on the surface with geodesic length\footnote{This relation assumes that the Gaussian curvature of the uniformized surface is $K = -1$.} $L_i = 4\pi p_i$. In the language of Fuchsian geometry, these singularities correspond to hyperbolic elements of the Fuchsian group.
    \item The case with $p_i = 0$ corresponds to a four-punctured Riemann sphere that was studied previously in \cite{Harrison:2022frl}. These singularities correspond to parabolic elements of the Fuchsian group.
    \item The case with $p_i \in \ic \bbR^+$ corresponds to the insertion of conical defects on the surface. In this case, $p_i^2 < 0$, and the operators correspond to the Fuchsian group's elliptic elements. For these singularities, we define $\kappa_i \defeq - \ic p_i$, which is related to the deficit angle $\theta$ of the conical defect by $\theta_i = 4 \pi \kappa_i \in [0, \pi)$.
\end{itemize}

It is, in principle, possible to consider a different value for each momentum or even consider general cases with mixed types of singularities, including boundaries, punctures, and conical defects; however, for simplicity, we focus on the case where all the external Liouville momenta are equal $p_1 = p_2 = p_3 = p_4 = p_{\circ}$ (or $\kappa_1 = \kappa_2 = \kappa_3 = \kappa_4 = \kappa_{\bullet}$ for conical defects). The conformal block decomposition of such a four-point function on the sphere is
\begin{equation}
    \expval{V_{\alpha_{\circ}}(\infty) V_{\alpha_{\circ}}(1) V_{\alpha_{\circ}}(x,\conj{x}) V_{\alpha_{\circ}}(0)} = \int_{\bbS_Q} \diff \beta \, C(\alpha_{\circ},\alpha_{\circ},\beta) C(Q-\beta,\alpha_{\circ},\alpha_{\circ}) \cF_{\circ}(\beta; x) \cF_{\circ}(\beta; \conj{x}),
\end{equation}
where $\cF_{\circ}(\beta;x)$ is the four-point Virasoro block corresponding to the insertion of four primaries with dimension $\Delta_{\circ} = \alpha_{\circ}(Q-\alpha_{\circ})$.

We first compute the semiclassical limit of the three-point structure coefficients and the conformal blocks. Performing a saddle-point analysis on the conformal block expansion allows us to compute the geodesic length of the cut, using which we compute the volumes of the moduli spaces of these surfaces and compare them to the volume polynomials. Following our convention in equation~(\ref{eq:ex:dimconv}), in the semiclassical limit,
\begin{equation}
    \lim_{b \to 0} \expval{V_{\alpha_{\circ}}(\infty) V_{\alpha_{\circ}}(1) V_{\alpha_{\circ}}(x,\conj{x}) V_{\alpha_{\circ}}(0)} \approx \int \diff p \, \exp(-b^{-2} S_{\circ}(p;x,\conj{x})),
\end{equation}
where
\begin{equation}
    S_{\circ}(p;x,\conj{x}) = S_3(p_{\circ},p_{\circ},p) + r(p) + S_3(p,p_{\circ},p_{\circ}) - f_{\circ}(p;x) - f_{\circ}(p;\conj{x}).
\end{equation}
The semiclassical blocks $f_{\circ}(p;x)$ that we consider in this setup correspond to the semiclassical limit of $\cF_{\circ}(\beta;x)$, where
\begin{equation}
    \lim_{b\to 0} b \beta = \frac{1}{2} + \ic p.
\end{equation}
One can obtain these blocks by taking the semiclassical limit of Zamolodchikov's recursion relations up to arbitrary order in the cross-ratio. In what follows, we will use $q$-expansion of the conformal blocks. The nome $q = \ep^{\ic \pi \tau}$ is related to the cross-ratio via equation~(\ref{eq:torusparameter}).

\subsubsection{Saddle-point analysis}
In our computations, we first assume that the four-point function corresponds to the insertion of four primary vertex operators in Liouville theory, i.e., $p_{\circ}^2 > 0$, corresponding to the case where the singularities are boundaries on the Riemann sphere. We then proceed to consider the case with $p_{\circ}^2 < 0$, which corresponds to the insertion of conical defects on the sphere by analytic continuation of the saddle-point equation in $p_{\circ} = \ic \kappa_{\bullet}$. In both cases, we assume all four external momenta are equal.

The total contribution of the three-point coefficients for identical external operators with momentum $p_{\circ}$ is
\begin{equation}
    \hat{S}_3(p) = S_3(p_{\circ},p_{\circ},p) + r(p) + S_3(p,p_{\circ},p_{\circ}),
\end{equation}
which can be computed using the formulas in \S\ref{a:semiclassics:dozz}. To perform the saddle-point analysis, we first compute the derivative of the three-point coefficient $\hat{S}_3(p)$,
\begin{equation}\label{res:sphere4geo:3ptderivative}
    \frac{\partial \hat{S}_3(p)}{\partial p} = 2\ic\log \frac{\Gamma^2\qty(1-2\ic p) \Gamma^2\qty(\frac{1}{2}+\ic p) \Gamma\qty(\frac{1}{2}+\ic p-2\ic p_{\circ}) \Gamma\qty(\frac{1}{2}+\ic p+2\ic p_{\circ})}{\Gamma^2\qty(1+2\ic p) \Gamma^2\qty(\frac{1}{2}-\ic p) \Gamma\qty(\frac{1}{2}-\ic p-2\ic p_{\circ}) \Gamma\qty(\frac{1}{2}-\ic p+2\ic p_{\circ})} - 2\pi,
\end{equation}
and then plug this into the saddle-point equation~(\ref{eq:pro:saddle}), which becomes,
\begin{equation}\label{eq:geodesicsaddle}
\ic\log \frac{\Gamma^2\qty(1-2\ic p) \Gamma^2\qty(\frac{1}{2}+\ic p) \Gamma\qty(\frac{1}{2}+\ic p-2\ic p_{\circ}) \Gamma\qty(\frac{1}{2}+\ic p+2\ic p_{\circ})}{\Gamma^2\qty(1+2\ic p) \Gamma^2\qty(\frac{1}{2}-\ic p) \Gamma\qty(\frac{1}{2}-\ic p-2\ic p_{\circ}) \Gamma\qty(\frac{1}{2}-\ic p+2\ic p_{\circ})} - \pi = \Re \frac{\partial f_{\circ}(p;q)}{\partial p},
\end{equation}
where the derivative of the classical block (see \S\ref{a:semiclassics:4pt}) up to $O(q^2)$ is
\begin{equation}\label{res:sphere4geo:blockderivative}
    \frac{\partial f_{\circ}(p;q)}{\partial p} = 2p \log 16q - \frac{p (1+16p_{\circ}^2)^2}{2(1+p^2)^2} q^2 + O\qty(q^4).
\end{equation}
In principle, the classical block and its derivative can be computed up to arbitrary orders in $q$-expansion using Zamolodchikov's recursion relations. 

Similarly, in the case of a sphere with conical defects, the saddle-point equation becomes,
\begin{equation}\label{eq:conicalsaddle}
    \ic\log \frac{\Gamma^2\qty(1-2\ic p) \Gamma^2\qty(\frac{1}{2}+\ic p) \Gamma\qty(\frac{1}{2}+\ic p+2\kappa_{\bullet}) \Gamma\qty(\frac{1}{2}+\ic p-2\kappa_{\bullet})}{\Gamma^2\qty(1+2\ic p) \Gamma^2\qty(\frac{1}{2}-\ic p) \Gamma\qty(\frac{1}{2}-\ic p+2\kappa_{\bullet}) \Gamma\qty(\frac{1}{2}-\ic p-2\kappa_{\bullet})} - \pi = \Re \frac{\partial f_{\bullet}(p;q)}{\partial p},
\end{equation}
where $f_{\bullet}(p;q)$ is obtained by replacing $p_{\circ}$ with $\ic \kappa_{\bullet}$ in the expression for the classical block $f_{\circ}(p;q)$. 

\begin{figure}
    \centering
    \input{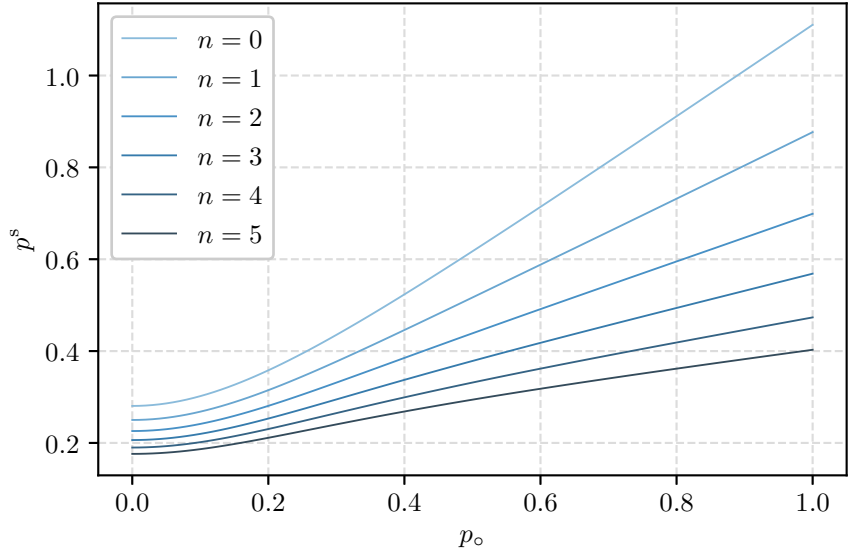}
    \caption{Saddle-point momentum $p^{\rms}$ for $q = \ep^{-\qty(1+\frac{n}{5})\pi}$ at $O\qty(q^2)$, sphere with four boundaries of length $L_{\circ} = 4\pi p_{\circ}$.}\label{fig:l04:allp:res}
\end{figure}

\begin{figure}
    \centering
    \input{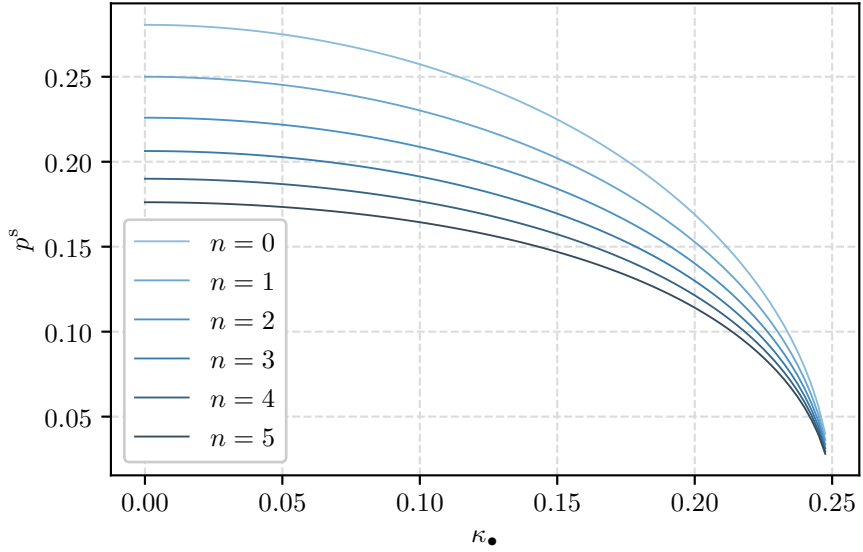}
    \caption{Saddle-point momentum $p^{\rms}$ for $q = \ep^{-\qty(1+\frac{n}{5})\pi}$ at $O\qty(q^2)$, sphere with four conical defects of angle $\theta_{\bullet} = 4\pi \kappa_{\bullet}$.}\label{fig:l04:allk:res}
\end{figure}

We numerically compute the saddle-point momenta $p^{\rms}$ which solve equations~(\ref{eq:geodesicsaddle}) and~(\ref{eq:conicalsaddle}) in the case of four external geodesic boundaries of momentum $p_\circ$ and cone points of momentum $\ic\kappa_\bullet$, respectively. We plot the results for $p^{\rms}$  for different values cross-ratio and of $p_{\circ}$ and $\kappa_{\bullet}$, including an expansion of the conformal block up to $O(q^2)$, in figures~\ref{fig:l04:allp:res} and~\ref{fig:l04:allk:res}, respectively.

\subsubsection{Comparison with exact results}
For particular values of the external momenta and cross-ratios, the Fuchsian construction of the corresponding hyperbolic surface is known explicitly. In these cases, we can compare our numerical results for the saddle-point momentum to the exact value obtained by taking the trace of appropriate elements of the corresponding Fuchsian group. Here, we would like to showcase two of these cases as examples.

\paragraph{Sphere with four punctures}
First, for the four-punctured sphere ($p_{\circ} = 0$ or equivalently $\kappa_{\bullet} = 0$) with cross-ratio $x = \frac{1}{2}$ corresponding to $q = \ep^{-\pi}$, the Fuchsian group is known explicitly \cite{Hempel:1988}. The four generators are
\begin{equation}
    \ug_1 = \begin{pmatrix}
        -1 & 0 \\
        2 & -1
    \end{pmatrix}, \quad
    \ug_2 = \begin{pmatrix}
        3 & -4 \\
        4 & -5
    \end{pmatrix}, \quad
    \ug_3 = \begin{pmatrix}
        3 & -8 \\
        2 & -5
    \end{pmatrix}, \quad
    \ug_4 = \begin{pmatrix}
        -1 & -4 \\
        0 & -1
    \end{pmatrix}.
\end{equation}
Using these generators, we can find the analytic value of the saddle-point momentum $p^{\rms}$ to be
\begin{equation}
    p^{\rms} = \frac{l}{4\pi} = \frac{1}{2\pi} \arccosh{\frac{\abs{\Tr \ug_1 \ug_2}}{2}} \approx 0.28055.
\end{equation}
In this case, which has been studied before in the literature \cite{Hadasz:2005gk,Harrison:2022frl}, the current method yields
\begin{equation}
    \begin{split}
        O\qty(q): \quad p^{\rms} &\approx 0.28057, \\
        O\qty(q^2): \quad p^{\rms} &\approx 0.28055, \\
        O\qty(q^4): \quad p^{\rms} &\approx 0.28055.
    \end{split}
\end{equation}
Here $O(q^n)$ refers to the order $n$ of the $q$-expansion for the classical blocks. Comparing the result with the analytic value, we find the error,
\begin{equation}
    \Delta_l \defeq \frac{\cosh(2\pi p^{\rms})}{\cosh \frac{l}{2}} - 1,
\end{equation}
to be
\begin{equation}
    \begin{split}
        O\qty(q): \quad \Delta_l &\approx 1.3 \times 10^{-4}, \\
        O\qty(q^2): \quad \Delta_l &\approx 1.2 \times 10^{-7}, \\
        O\qty(q^4): \quad \Delta_l &\approx 1.1 \times 10^{-11}.
    \end{split}
\end{equation}

\paragraph{Sphere with four conical defects}
Now we consider the sphere with four conical singularities, each with a deficit angle of $\frac{2\pi}{3}$. This particular surface is intricately connected to a genus two construct, which can be realized through the following algebraic curve equation:
\begin{equation}\label{eq:ex:gentwo:alg}
    y^2 = z^6 - 1,
\end{equation}
with a corresponding period matrix \cite{Kuusalo:1995ge} given by
\begin{equation}\label{eq:ex:conic:period}
    \Omega = \frac{\ic}{\sqrt{3}} \begin{pmatrix}
        2 & -1 \\
        -1 & 2
    \end{pmatrix}.
\end{equation}
The Fuchsian generators for this genus two surface are
\begin{equation}\label{eq:ex:conic:gentwo}
    T_k = \begin{pmatrix}
        1 + 2 \conj{\omega} & - \omega^{k-1} \sqrt{6} \\
        - \conj{\omega}^{k-1} \sqrt{6} & 1 + 2 \omega
    \end{pmatrix}, \quad w = \ep^{\frac{\ic\pi}{3}}, \quad k \in {1,2,4,5},
\end{equation}
satisfying
\begin{equation}
    T_4^{-1} T_5 T_4 T_5^{-1} T_1^{-1} T_2 T_1 T_2^{-1} = 1,
\end{equation}
which is the usual construction of a genus two surface by gluing the sides of an octagon together. This also means that the collection $\{T_1, T_2, T_4, T_5\}$ correspond to the canonical $\alpha$ and $\beta$ cycles of the genus two surface in equation~(\ref{eq:ex:gentwo:alg}). Each of the $T_i$ constructed here satisfies $\abs{\Tr T_i} = 4$, which means that the length of each cycle is $l = 2 \arccosh{2}$.

The same genus two surface can be represented as a three-fold cover of the Riemann sphere defined as
\begin{equation}
    y^3 = \frac{z(z-1)}{z - \frac{1}{2}},
\end{equation}
which exhibits four conical defects with a deficit angle of $\theta_{\bullet} = \frac{2\pi}{3}$ (or equivalently, $\kappa_{\bullet} = \frac{1}{6}$) at cross-ratio $x = \frac{1}{2}$. This surface has the same period matrix as equation~(\ref{eq:ex:conic:period}). In this representation, a Fuchsian generator $T_i$ in equation~(\ref{eq:ex:conic:gentwo}) can be thought of as a cycle enclosing two of the conical singularities. Hence, the saddle-point momentum $p^{\rms}$ at cross-ratio $x = \frac{1}{2}$ for the case with four conical singularities with deficit angle $\theta_{\bullet} = \frac{2\pi}{3}$ should be compared to
\begin{equation}\label{eq:ex:con:asol}
    p^{\rms} = \frac{l}{4\pi} = \frac{1}{2\pi} \arccosh \frac{\abs{\Tr T_i}}{2} \approx 0.209600.
\end{equation}
Indeed, by performing the saddle-point analysis, we find the result to be
\begin{equation}
    \begin{split}
        O\qty(q): \quad p^{\rms} &\approx 0.209605, \\
        O\qty(q^2): \quad p^{\rms} &\approx 0.209600, \\
        O\qty(q^4): \quad p^{\rms} &\approx 0.209600,
    \end{split}
\end{equation}
and the corresponding errors to be
\begin{equation}
    \begin{split}
        O\qty(q): \quad \Delta_l &\approx 2.4 \times 10^{-5}, \\
        O\qty(q^2): \quad \Delta_l &\approx 3.6 \times 10^{-8}, \\
        O\qty(q^4): \quad \Delta_l &\approx 3.6 \times 10^{-11}.
    \end{split}
\end{equation}

\begin{figure}
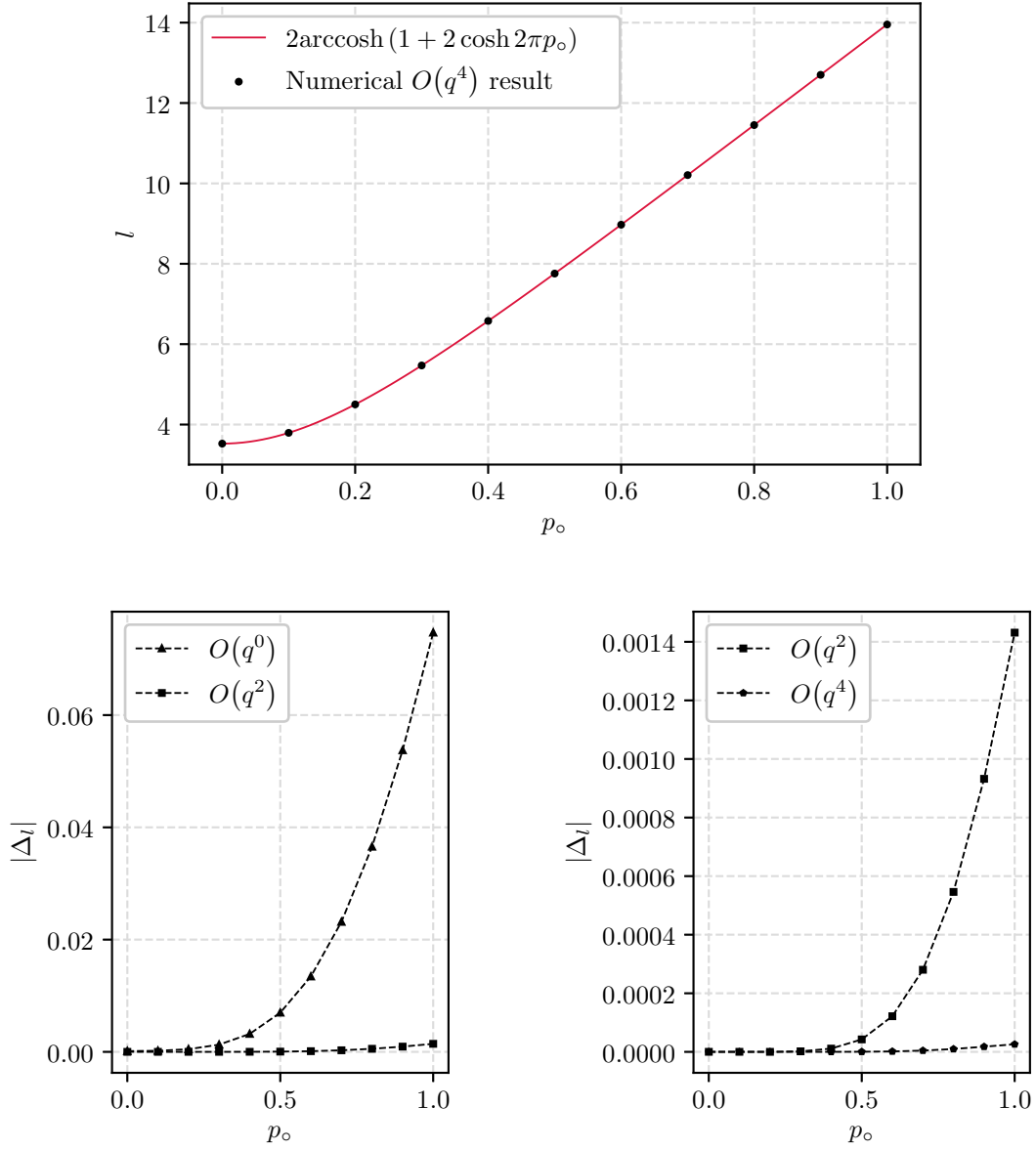

    \centering
    \begin{subfigure}{\textwidth}
        \centering
        \input{geolencomp.pgf}
    \end{subfigure}\vspace{1em}
    \begin{subfigure}{.47\textwidth}
        \centering
        \input{geolener1.pgf}
    \end{subfigure}
    \begin{subfigure}{.47\textwidth}
        \centering
        \input{geolener2.pgf}
    \end{subfigure}
    \caption{Comparison between the numerical results for the saddle-point length at cross-ratio $x = \frac{1}{2}$ and the formula in equation~(\ref{eq:l04:form}). The external operators all correspond to geodesic boundaries of length $L_\circ = 4 \pi p_\circ$.}\label{fig:l04:allp:comp}
\end{figure}

\begin{figure}
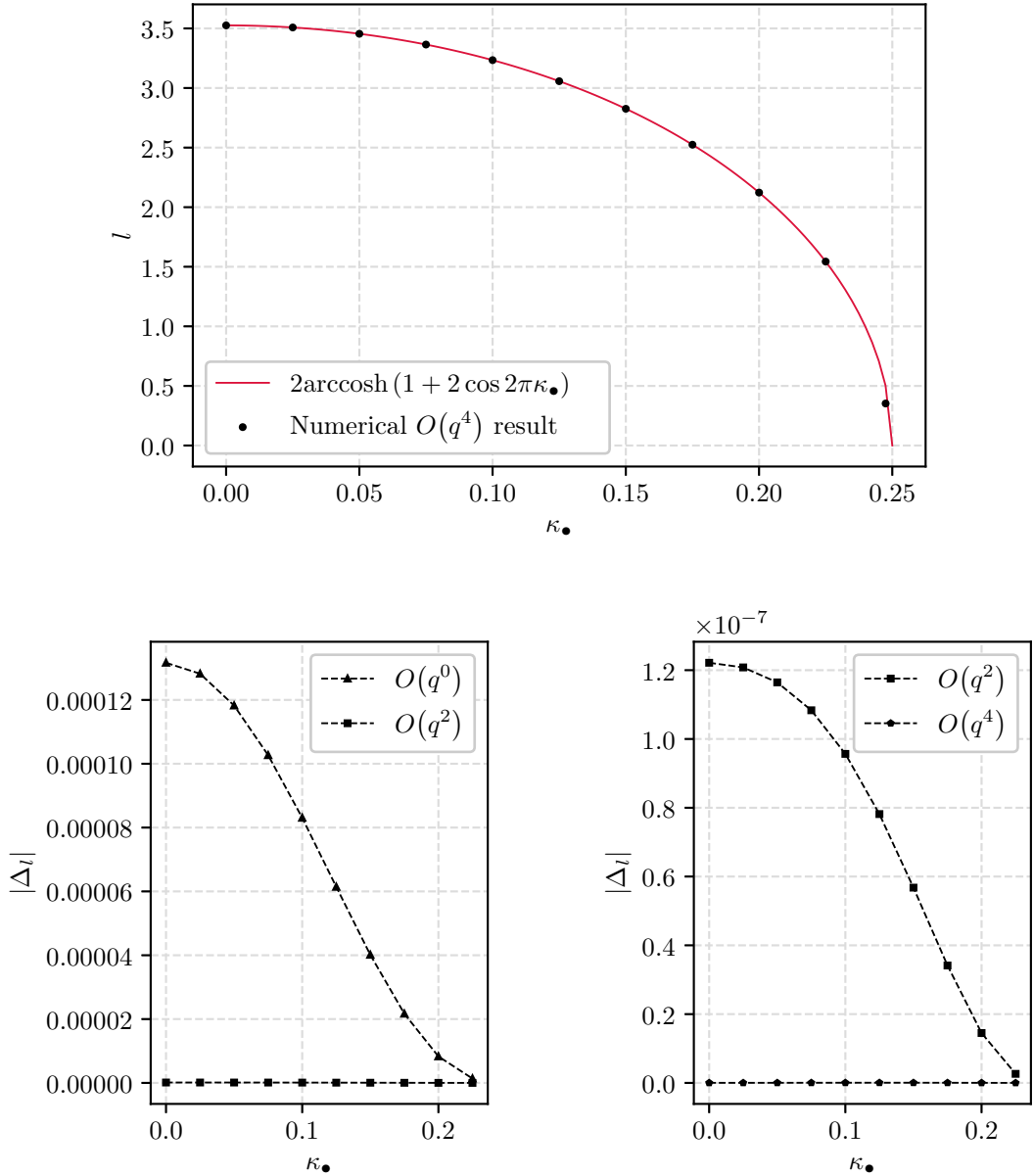

    \centering
    \begin{subfigure}{\textwidth}
        \centering
        \input{conlencomp.pgf}
    \end{subfigure}\vspace{1em}
    \begin{subfigure}{.50\textwidth}
        \centering
        \input{conlener1.pgf}
    \end{subfigure}
    \begin{subfigure}{.44\textwidth}
        \centering
        \input{conlener2.pgf}
    \end{subfigure}
    \caption{Comparison between the numerical results for the saddle-point length at cross-ratio $x = \frac{1}{2}$ and the formula in equation~(\ref{eq:l04:form}). The external operators all correspond to conical defects with angle $\theta_\bullet = 4 \pi \kappa_\bullet$.}\label{fig:l04:allk:comp}
\end{figure}

In fact, this analysis can be generalized to an arbitrary surface with four conical defects (or four boundaries) and cross-ratio $x = \frac{1}{2}$. 
The value of the geodesic length $l^{\rms}$ is given by
\begin{equation}\label{eq:l04:form}
    \cosh \frac{l}{2} = 1 + 2 \cosh \frac{l_{\circ}}{2},
\end{equation}
for surfaces with four boundaries of length $l_{\circ}$, and
\begin{equation}
    \cosh \frac{l}{2} = 1 + 2 \cos \frac{\theta_{\bullet}}{2},
\end{equation}
for surfaces with four conical defects of deficit angle $\theta_{\bullet}$. Again, our numerics provide excellent agreement with these analytic formulae. The comparison between these formulae and the numerical results are shown in figures~\ref{fig:l04:allp:comp} and~\ref{fig:l04:allk:comp}.

\subsubsection{Volume computation}
\begin{figure}
    \centering
    \input{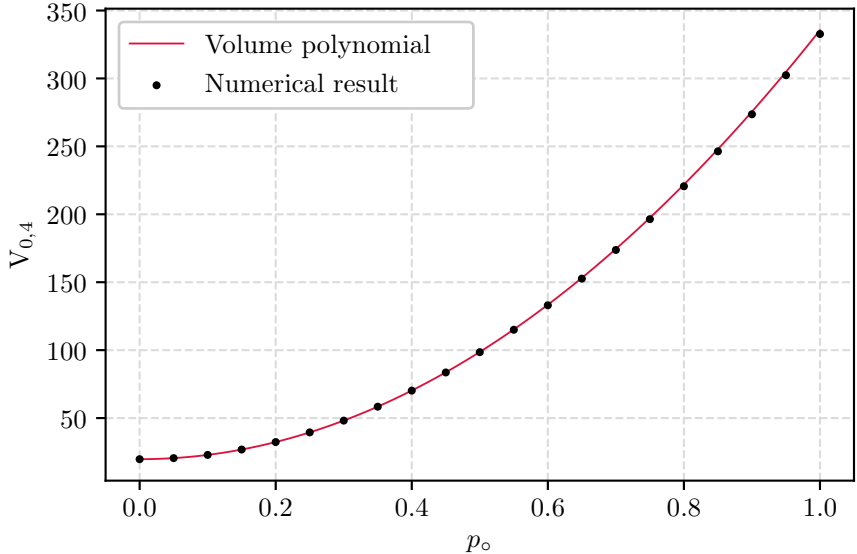}
    \caption{Volumes of moduli space at $O\qty(q)$, sphere with four boundaries, $L_1 = L_2 = L_3 = L_4 = 4 \pi p_{\circ}$.}\label{fig:v04:allp:res}
\end{figure}

\begin{figure}
    \centering
    \input{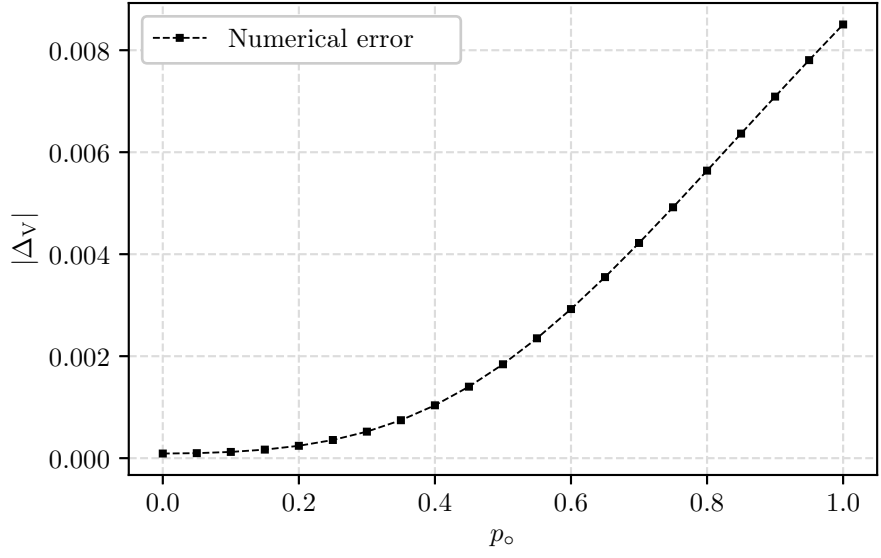}
    \caption{Errors for the volumes at $O\qty(q)$, sphere with four boundaries, $L_1 = L_2 = L_3 = L_4 = 4 \pi p_{\circ}$.}\label{fig:v04:allp:err}
\end{figure}

\begin{figure}
    \centering
    \input{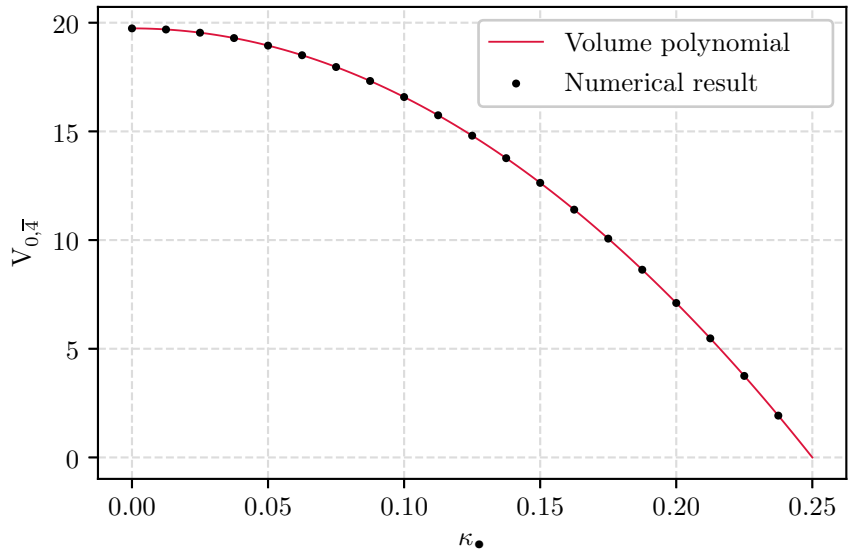}
    \caption{Volumes of moduli space at $O\qty(q)$, sphere with four conical defects, $\theta_1 = \theta_2 = \theta_3 = \theta_4 = 4\pi \kappa_{\bullet}$.}\label{fig:v04:allk:res}
\end{figure}

\begin{figure}
    \centering
    \input{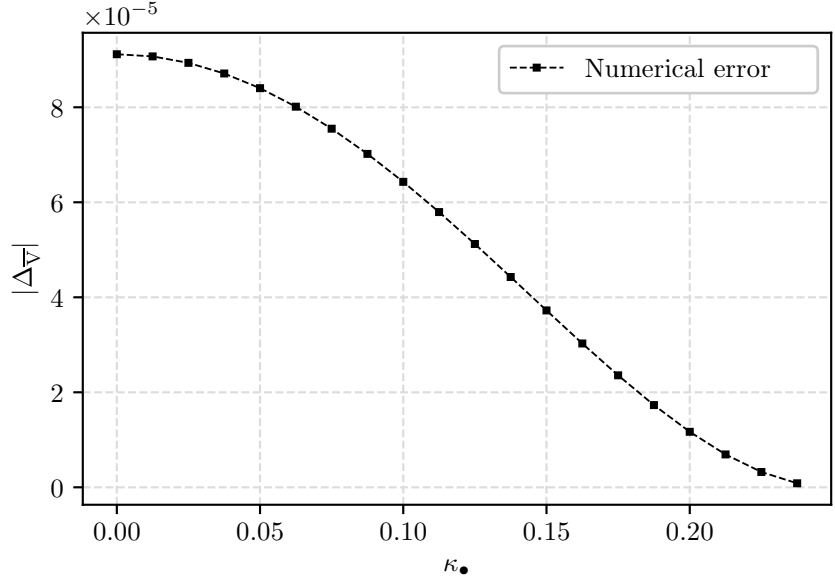}
    \caption{Errors for the volumes at $O\qty(q)$, sphere with four conical defects, $\theta_1 = \theta_2 = \theta_3 = \theta_4 = 4\pi \kappa_{\bullet}$.}\label{fig:v04:allk:err}
\end{figure}

Now that we have found the saddle-point momenta, we can use equation~(\ref{eq:pro:metricfromps}) to compute the Weil--Petersson metric $g_{\WP}$ and integrate $\sqrt{g_{\WP}}$ over the moduli space to compute the Weil--Petersson volume of a sphere with four singularities. We denote the result of our computations for the volume of the moduli space of a sphere with four geodesic boundaries as $\rV_{0,4}$ and for the volume of the moduli space of a sphere with conical defects as $\rV_{0,\conj{4}}$. Our results for these volumes are plotted in figures~\ref{fig:v04:allp:res} and~\ref{fig:v04:allk:res}. Compared to the volume polynomials $\sV_{0,4}$ and $\sV_{0,\conj{4}}$ in equations~(\ref{eq:4bpolynomial}) and~(\ref{eq:4cpolynomial}) respectively, we define the errors as
\begin{equation}
    \Delta_{\rV} \defeq \frac{\rV_{0,4}(\boldsymbol{L})}{\sV_{0,4}(\boldsymbol{L})} - 1, \quad \Delta_{\conj{\rV}} \defeq \frac{\rV_{0,\conj{4}}(\boldsymbol{\theta})}{\sV_{0,\conj{4}}(\boldsymbol{\theta})} - 1
\end{equation}
and present them in figures~\ref{fig:v04:allp:err} and~\ref{fig:v04:allk:err} for bordered and cone surfaces respectively.

\subsection{\texorpdfstring{$\bbZ_3$-symmetric genus two surfaces}{Z3-symmetric genus two surfaces}}\label{s:genus2Z3}
As another example, we focus on the case of $\bbZ_3$-symmetric genus two Riemann surfaces $\rY_{3,2}$ described in \S\ref{s:algcurves} as the algebraic curve
\begin{equation}
    y^3 = \frac{z(z-1)}{z - x},
\end{equation}
with the period matrix given in equation~(\ref{eq:y32periodmatrix}). We call the variable $x$ cross-ratio since it corresponds to the cross-ratio of four twist-3 operators on the Riemann sphere in the language of symmetric product orbifold CFTs.

As previously mentioned in \S\ref{s:algcurves}, this family of surfaces has a one-complex-dimensional moduli space with coordinates $(x,\conj{x})$. Alternatively, we can use the coordinates $(q,\conj{q})$ related to the modular parameter $\tau$ defined in equation~(\ref{eq:torusparameter}) via $q = \ep^{\ic \pi \tau}$ as the coordinate on moduli space.

The partition function of Liouville theory on these surfaces can be decomposed in terms of conformal blocks as
\begin{equation}
    \cZ^{\rY_{3,2}}(q,\conj{q}) = \int_{\bbS_Q^3} \diff \boldsymbol{\beta} \, C(\beta_1,\beta_2,\beta_3) C(Q-\beta_1,Q-\beta_2,Q-\beta_3) \cF^{\rY_{3,2}}(\boldsymbol{\beta};q) \cF^{\rY_{3,2}} (\boldsymbol{\beta};\conj{q}),
\end{equation}
where $\cF^{\rY_{3,2}}$ are the conformal blocks on $\bbZ_3$-symmetric genus two surfaces which can be found in \S\ref{a:twistblocks}.\footnote{In \S\ref{a:twistblocks} we illustrate a way to calculate the conformal blocks for $\bbZ_3$-symmetric R\'{e}nyi surfaces as an expansion in cross-ratio $x$. At higher orders, we use the results of \cite{Cho:2017fzo} and thank the authors for providing their code for the computation of these blocks.} In the semiclassical limit,
\begin{equation}
    \lim_{b\to 0} \cZ^{\rY_{3,2}}(q,\conj{q}) \approx \int \diff p_1 \diff p_2 \diff p_3 \, \exp(-b^{-2} S^{\rY_{3,2}}(p_1,p_2,p_3;q,\conj{q})),
\end{equation}
where
\begin{equation}
    \begin{split}
        S^{\rY_{3,2}}(p_1,p_2,p_3;q,\conj{q}) &= 2 S_3(p_1,p_2,p_3) + r(p_1) + r(p_2) + r(p_3) \\
        &\quad- f^{\rY_{3,2}}(p_1,p_2,p_3;q) - f^{\rY_{3,2}}(p_1,p_2,p_3;\conj{q}).
    \end{split}
\end{equation}
Here, $f^{\rY_{3,2}}(p_1,p_2,p_3;q)$ are the semiclassical blocks corresponding to the $\bbZ_3$-symmetric genus two conformal blocks with
\begin{equation}
    \lim_{b \to 0} b \beta_i = \frac{1}{2} + \ic p_i.
\end{equation}

\subsubsection{Saddle-point analysis}
Similar to the previous example, in order to solve the saddle-point equations, we need to compute the derivatives of the three-point coefficients and the conformal blocks with respect to the exchange momenta. The total contribution of three-point coefficients is
\begin{equation}
    \hat{S}_3^{\rY_{3,2}}(p_1,p_2,p_3) = 2 S_3(p_1,p_2,p_3) + r(p_1) + r(p_2) + r(p_3).
\end{equation}
The derivatives of the total contribution of three-point coefficients $\hat{S}_3^{\rY_{3,2}}(p_1,p_2,p_3)$ with respect to the internal momenta $p_1$, $p_2$, and $p_3$ are
\begin{align}
    \frac{\partial \hat{S}_3^{\rY_{3,2}}(p_1,p_2,p_3)}{\partial p_1} &= 4\ic\log \frac{\Gamma\qty(1-2\ic p_1)}{\Gamma\qty(1+2\ic p_1)} + 2 \ic \sum_{\sigma_i = \pm 1} \log \frac{\Gamma\qty(\frac{1}{2} + \ic p_1 + \ic \sigma_2 p_2 + \ic \sigma_3 p_3)}{\Gamma\qty(\frac{1}{2} - \ic p_1 - \ic \sigma_2 p_2 - \ic \sigma_3 p_3)} - 2\pi, \\
    \frac{\partial \hat{S}_3^{\rY_{3,2}}(p_1,p_2,p_3)}{\partial p_2} &= 4\ic\log \frac{\Gamma\qty(1-2\ic p_2)}{\Gamma\qty(1+2\ic p_2)} + 2 \ic \sum_{\sigma_i = \pm 1} \sigma_2 \log \frac{\Gamma\qty(\frac{1}{2} + \ic p_1 + \ic \sigma_2 p_2 + \ic \sigma_3 p_3)}{\Gamma\qty(\frac{1}{2} - \ic p_1 - \ic \sigma_2 p_2 - \ic \sigma_3 p_3)} - 2\pi, \\
    \frac{\partial \hat{S}_3^{\rY_{3,2}}(p_1,p_2,p_3)}{\partial p_3} &= 4\ic\log \frac{\Gamma\qty(1-2\ic p_3)}{\Gamma\qty(1+2\ic p_3)} + 2 \ic \sum_{\sigma_i = \pm 1} \sigma_3 \log \frac{\Gamma\qty(\frac{1}{2} + \ic p_1 + \ic \sigma_2 p_2 + \ic \sigma_3 p_3)}{\Gamma\qty(\frac{1}{2} - \ic p_1 - \ic \sigma_2 p_2 - \ic \sigma_3 p_3)} - 2\pi.
\end{align}

The saddle-point equations for this case are then
\begin{align}
    \label{eq:ex:gen2:sadeqs}
    2 \ic\log \frac{\Gamma\qty(1-2\ic p_1)}{\Gamma\qty(1+2\ic p_1)} + \ic \sum_{\sigma_i = \pm 1} \log \frac{\Gamma\qty(\frac{1}{2} + \ic p_1 + \ic \sigma_2 p_2 + \ic \sigma_3 p_3)}{\Gamma\qty(\frac{1}{2} - \ic p_1 - \ic \sigma_2 p_2 - \ic \sigma_3 p_3)} - \pi &= \Re \frac{\partial f^{\rY_{3,2}}(p_1,p_2,p_3;q)}{\partial p_1}, \\
    2 \ic\log \frac{\Gamma\qty(1-2\ic p_2)}{\Gamma\qty(1+2\ic p_2)} + \ic \sum_{\sigma_i = \pm 1} \sigma_2 \log \frac{\Gamma\qty(\frac{1}{2} + \ic p_1 + \ic \sigma_2 p_2 + \ic \sigma_3 p_3)}{\Gamma\qty(\frac{1}{2} - \ic p_1 - \ic \sigma_2 p_2 - \ic \sigma_3 p_3)} - \pi &= \Re \frac{\partial f^{\rY_{3,2}}(p_1,p_2,p_3;q)}{\partial p_2}, \\
    2 \ic\log \frac{\Gamma\qty(1-2\ic p_3)}{\Gamma\qty(1+2\ic p_3)} + \ic \sum_{\sigma_i = \pm 1} \sigma_3 \log \frac{\Gamma\qty(\frac{1}{2} + \ic p_1 + \ic \sigma_2 p_2 + \ic \sigma_3 p_3)}{\Gamma\qty(\frac{1}{2} - \ic p_1 - \ic \sigma_2 p_2 - \ic \sigma_3 p_3)} - \pi &= \Re \frac{\partial f^{\rY_{3,2}}(p_1,p_2,p_3;q)}{\partial p_3},
\end{align}
where the derivatives of the conformal blocks $f^{\rY_{3,2}}(p_1,p_2,p_3;q)$ (see \S\ref{a:semiclassics:gen2}) with respect to the momenta $p_i$ up to $O(q)$ are
\begin{align}
    \frac{\partial f^{\rY_{3,2}}(p_1,p_2,p_3;q)}{\partial p_1} &= 2p_1 \log \frac{16q}{27} + \frac{128}{27} p_1 \qty(\frac{p_1^2 - p_2^2}{1+4p_3^2} + \frac{p_1^2 - p_3^2}{1+4p_2^2} - 2 \frac{(p_2^2 - p_3^2)^2}{(1+4p_1^2)^2}) q + O\qty(q^2), \\
    \frac{\partial f^{\rY_{3,2}}(p_1,p_2,p_3;q)}{\partial p_2} &= 2p_2 \log \frac{16q}{27} + \frac{128}{27} p_2 \qty(\frac{p_2^2 - p_3^2}{1+4p_1^2} +\frac{p_2^2 - p_1^2}{1+4p_3^2} - 2 \frac{(p_3^2 - p_1^2)^2}{(1+4p_2^2)^2}) q + O\qty(q^2), \\
    \frac{\partial f^{\rY_{3,2}}(p_1,p_2,p_3;q)}{\partial p_3} &= 2p_3 \log \frac{16q}{27} + \frac{128}{27} p_3 \qty(\frac{p_3^2 - p_1^2}{1+4p_2^2} + \frac{p_3^2 - p_2^2}{1+4p_1^2} - 2 \frac{(p_1^2 - p_2^2)^2}{(1+4p_3^2)^2}) q + O\qty(q^2).
\end{align}

Since the conformal blocks are invariant under permuting the exchange momenta, by solving equation~(\ref{eq:ex:gen2:sadeqs}), we inevitably fall into the locus where $p_1 = p_2 = p_3$. Hence, we can treat the momenta as equal from the beginning and use
\begin{equation}
    \frac{\partial \hat{S}_3^{\rY_{3,2}}(p,p,p)}{\partial p} = 6 \ic\log \frac{\Gamma^2\qty(1-2\ic p) \Gamma\qty(\frac{1}{2}+\ic p) \Gamma\qty(\frac{1}{2}+3\ic p)}{\Gamma^2\qty(1+2\ic p) \Gamma\qty(\frac{1}{2}-\ic p) \Gamma\qty(\frac{1}{2}-3\ic p)} - 6\pi.
\end{equation}
In this case, the saddle-point equation becomes
\begin{equation}\label{eq:ex:gen2:sadeq}
    3\ic\log \frac{\Gamma^2\qty(1-2\ic p) \Gamma\qty(\frac{1}{2}+\ic p) \Gamma\qty(\frac{1}{2}+3\ic p)}{\Gamma^2\qty(1+2\ic p) \Gamma\qty(\frac{1}{2}-\ic p) \Gamma\qty(\frac{1}{2}-3\ic p)} - 3\pi= \Re \frac{\partial f^{\rY_{3,2}}(p,p,p;q)}{\partial p},
\end{equation}
where the derivative of the classical block $\partial f(p,p,p;q)$ with equal exchange momenta with respect to $p$ is
\begin{equation}
    \frac{\partial f^{\rY_{3,2}}(p,p,p;q)}{\partial p} = 6 p \log \frac{16q}{27} - \frac{25 p}{54 (1+p^2)^2} q^2 + O\qty(q^4).
\end{equation}

The saddle-point equation~(\ref{eq:ex:gen2:sadeq}) is solved numerically to yield the following saddle-point momentum:
\begin{equation}
    \begin{split}
        O\qty(q): \quad p^{\rms} &\approx 0.209605, \\
        O\qty(q^2): \quad p^{\rms} &\approx 0.209600, \\
        O\qty(q^4): \quad p^{\rms} &\approx 0.209600,
    \end{split}
\end{equation}
for $q = \ep^{-\pi}$ corresponding to cross-ratio value $x = \frac{1}{2}$. For these surfaces, we should compare our results to the analytic solution we mentioned in equation~(\ref{eq:ex:con:asol}). We define the numerical error to be
\begin{equation}
    \Delta_l \defeq \frac{\cosh(2\pi p^{\rms})}{\cosh \frac{l}{2}} - 1
\end{equation}
and find that our numerical result differs from the exact value by
\begin{align}
    O\qty(q): \quad \Delta_l &\approx 2.4 \times 10^{-5}, \\
    O\qty(q^2): \quad \Delta_l &\approx 3.6 \times 10^{-8}, \\
    O\qty(q^4): \quad \Delta_l &\approx 3.6 \times 10^{-11}.
\end{align}

\subsubsection{Volume computation}
Having found the saddle-point momenta as a solution to the saddle-point equation, again, we can use the metric in equation~(\ref{eq:pro:metricfromps}) and integrate it over the one-complex-dimensional moduli space to compute the Weil--Petersson volume corresponding to the $\bbZ_3$-symmetric genus two surfaces $\rY_{3,2}$. The numerical results for the volume of the $\bbZ_3$-symmetric genus two surfaces at $O\qty(q)$ is
\begin{equation}
    \begin{split}
        O\qty(q): \quad \rV_{\rY_{3,2}} \approx 32.8978, \\
        O\qty(q^2): \quad \rV_{\rY_{3,2}} \approx 32.8987, \\
        O\qty(q^4): \quad \rV_{\rY_{3,2}} \approx 32.8987.
    \end{split}
\end{equation}
We can see that this volume is related to the volume of the moduli space of a sphere with four conical singularities, each having an angle of $\frac{\pi}{3}$. The precise numerical value for the volume of the moduli space of a sphere with four conical singularities with deficit angles of $\theta = \frac{\pi}{3}$ according to equation~(\ref{eq:4cpolynomial}) is
\begin{equation}
    \sV_{0,\conj{4}}\qty(\frac{\pi}{3}) = \frac{10}{9} \pi^2 \approx 10.9662.
\end{equation}
In this case, we define the numerical error as
\begin{equation}
    \Delta_{\rV} \defeq \frac{1}{3} \frac{\rV_{\rY_{3,2}}}{\sV_{0,\conj{4}}\qty(\frac{\pi}{3})} - 1,
\end{equation}
and find the error for our results to be
\begin{align}
    O\qty(q): \quad \Delta_\rV &\approx - 2.8 \times 10^{-5}, \\
    O\qty(q^2): \quad \Delta_\rV &\approx 5.4 \times 10^{-8}, \\
    O\qty(q^4): \quad \Delta_\rV &\approx - 2.9 \times 10^{-9}.
\end{align}
It is evident that the volume of the moduli space of the genus two R\'{e}nyi surface $\sV_{\rY_{3,2}}$ is three times the volume $\sV_{0,\conj{4}}\qty(\frac{\pi}{3})$ of the moduli space of the sphere with four conical defects of angle $\frac{\pi}{3}$. Moreover, the the Weil--Petersson metric on the moduli space of $\rY_{3,2}$ is three times the metric on $\cM_{0,\conj{4}}\qty(\frac{\pi}{3})$. This is due to the fact that the CFT living on $\rY_{3,2}$ is equivalent to three copies of the one living on $\Sigma_{0,\conj{4}}\qty(\frac{\pi}{3})$ and the central charges differ by a factor of three.

\section*{Acknowledgements}
We thank  S. Collier and T. Numasawa for helpful discussions.  
The work of S.M.H. was partially supported by the National Science and Engineering Council of Canada and the Canada Research Chairs program.
Research of A.M. is supported in part by the Simons Foundation Grant No. 385602 and the Natural Sciences and Engineering Research Council of Canada (NSERC), funding reference number
SAPIN/00047-2020.

\appendix
\section{Direct computation of genus two conformal blocks}\label{a:twistblocks}
In this appendix, we will calculate the conformal blocks for the $\bbZ_3$-symmetric genus two surface $\rY_{3,2}$. A similar computation comparing the torus partition function to the four-point function of twist two operators has been completed in \cite{Lunin:2000yv, Headrick:2010zt}. The computation here is performed by examining the four-point function of twist three operators $\sigma$ and $\conj{\sigma}$. Using the definition of the blocks in \S\ref{s:genus2Z3}, the four-point function is decomposed as
\begin{equation}
    \expval{\conj{\sigma}(\infty) \sigma(1) \conj{\sigma}(x) \sigma(0)} = \sum_{i,j,k} C_{ijk}^2 \cF^{\rY_{3,2}}(h_i, h_j, h_k; x) \cF^{\rY_{3,2}}(\conj{h}_i, \conj{h}_j, \conj{h}_k; \conj{x}),
\end{equation}
where the sum is over operators $\cO_i$, $\cO_j$, $\cO_k$ of our CFT, which we will denote as $\cC$. This decomposition can be compared to the usual decomposition of a four-point function into conformal blocks,
\begin{equation}
    \expval{\conj{\sigma}(\infty) \sigma(1) \conj{\sigma}(x) \sigma(0)} = \sum_{\tilde{\cO}} C_{\sigma \sigma \tilde{\cO}}^2 \cF_{\sigma \sigma}^{\sigma \sigma}(h_{\tilde{\cO}}; x)\cF_{\sigma \sigma}^{\sigma \sigma}(\conj{h}_{\tilde{\cO}}; \conj{x}),
\end{equation}
where the sum is over conformal primaries $\tilde{\cO}$ of the orbifold theory $\cC^3/\bbZ_3$. The conformal primaries in the orbifold theory are constructed from the primaries of the seed theory, which allows us to compare the two decompositions to obtain the relation
\begin{equation}
    C_{ijk}^2 \cF^{\rY_{3,2}}(h_i, h_j, h_k; x) \cF^{\rY_{3,2}}(\conj{h}_i, \conj{h}_j, \conj{h}_k; \conj{x}) = \sum_{\tilde{\cO}_{ijk}} C_{\sigma \sigma \tilde{\cO}_{ijk}}^2 \cF_{\sigma \sigma}^{\sigma \sigma}(h_{ijk}; x)\cF_{\sigma \sigma}^{\sigma \sigma}(\conj{h}_{ijk}; \conj{x}),
\end{equation}
where the sum is over all primaries $\tilde{\cO}_{ijk}$ of $\cC^3/\bbZ_3$ that are constructed from the primaries $\cO_i$, $\cO_j$, $\cO_k$ of the seed theory and their descendants.

The first operator in the orbifold theory that contributes involves only primary states of the seed operators,
\begin{equation}
    \tilde{\cO} = \frac{1}{\sqrt{3}} \qty(\cO_1 \otimes \cO_2 \otimes \cO_3 + \text{cyclic}),
\end{equation}
with conformal dimensions $\tilde{h} = h_1 + h_2 + h_3$, $\tilde{\conj{h}} = \conj{h}_1 + \conj{h}_2 + \conj{h}_3$. Here we are assuming the normalization
$$\expval{\cO_i (1,1) \cO_j(0,0)} = \delta_{ij}.$$ The factor of $\sqrt{3}$ above ensures that $\tilde{\cO}$ is similarly normalized.

The three-point coefficient $C_{\sigma \sigma \tilde{O}}$ can be calculated directly 
\begin{align}
    C_{\sigma \sigma \tilde{O}} &= \expval{\conj{\sigma}(\infty, \infty) \tilde{\cO}(z=1, \conj{z}=1) \sigma(0,0)} \\
    &= \frac{1}{\sqrt{3}} \biggl\langle\qty[(3t^2)^{-h_1}(3\conj{t}^2)^{-\conj{h}_1} \cO_1(t, \conj{t})]_{\subalign{t&=1 \\ \conj{t}&=1}} \qty[(3t^2)^{-h_2} (3\conj{t}^2)^{-\conj{h}_2} \cO_2(t, \conj{t})]_{\subalign{t&=\ep^{2\pi \ic/3} \\ \conj{t}&=\ep^{-2\pi \ic/3}}} \\ \nonumber
    & \qquad \qquad \qquad \times \qty[(3t^2)^{-h_3}(3\conj{t}^2)^{-\conj{h}_3} \cO_3(t, \conj{t})]_{\subalign{t&=\ep^{-2\pi \ic/3} \\ \conj{t}&=\ep^{2\pi \ic/3}}} + \text{cyclic} \biggr\rangle \\
    &= 3^{\frac{1}{2}-\frac{3}{2}(\Delta_1 + \Delta_2 + \Delta_3)} \ic^{\ell_1 + \ell_2 + \ell_3} C_{123},
\end{align}
where the second line follows from unwrapping the coefficient with the map $z = t^3$ and results in a sum of three-point functions involving the three seed operators. In these equations we use $\Delta_i = h_i + \conj{h}_i$ and $\ell_i = h_i - \conj{h}_i$ to denote the dimension and spin of the seed operators. Note that in the final line, we have the combination $\ic^{\ell_1 + \ell_2 + \ell_3} C_{123}$. As pointed out in \cite{Collier:2019weq}, $C_{123}$ is purely imaginary when $\ell_1 + \ell_2 + \ell_3$ is odd. The result is that the three-point coefficient $C_{\sigma \sigma \tilde{O}}$ is real.

New primaries can be constructed out of the descendants of $\cO_1$, $\cO_2$, and $\cO_3$. These new primaries must still respect the cyclic symmetry. We will focus our calculation on primaries of conformal dimensions $\tilde{h}+1$ and $\tilde{\conj{h}}$. This leads to the following building blocks,
\begin{align}
    \tilde{A}_1 &= \partial\cO_1 \otimes \cO_2 \otimes \cO_3 + \text{cyclic}, \\
    \tilde{A}_2 &= \cO_1 \otimes \partial \cO_2 \otimes \cO_3 + \text{cyclic}, \\
    \tilde{A}_3 &= \cO_1 \otimes \cO_2 \otimes \partial \cO_3 + \text{cyclic}.
\end{align}
One combination of these building blocks is simply $\partial \tilde{\cO}$ which is given by
\begin{equation}
    \partial \tilde{\cO} = \frac{1}{\sqrt{3}}(\tilde{A}_1 + \tilde{A}_2 + \tilde{A}_3).
\end{equation}
This leaves us with two new primaries at this level. The condition for a new operator
\begin{equation}
    \tilde{\cO}_{\text{new}} = \sum_{i=1}^3 \alpha_i \tilde{A}_i 
\end{equation}
to be primary is simply expressed as
\begin{equation}
    \sum_{i=1}^3 \alpha_i h_i = 0.
\end{equation}
Combining this with the requirement that the two new primaries be orthonormal with respect to the two-point function, we can take
\begin{equation}
    \begin{split}
    \tilde{\cO}_{\text{new}_1}&: \quad \alpha_1 = \frac{\ic}{\sqrt{6}} \frac{h_3}{\sqrt{h_1 h_3(h_1 + h_3)}}, \quad \alpha_2 = 0, \quad \alpha_3 = -\frac{\ic}{\sqrt{6}} \frac{h_1}{\sqrt{h_1 h_3 (h_1 + h_3)}}, \\
    \tilde{\cO}_{\text{new}_2}&: \quad \alpha_1 = \alpha_3 = -\frac{\ic}{\sqrt{6}} \frac{h_2}{\sqrt{h_2(h_1 + h_3)(h_1 + h_2 + h_3)}}, \quad \alpha_2 = \frac{\ic}{\sqrt{6}} \sqrt{\frac{h_1 + h_3}{h_2(h_1 + h_3 + h_2)}}.
    \end{split}
\end{equation}

The relevant three-point coefficients $C_{\sigma \sigma \tilde{\cO}_{\text{new}_1}}$ and $C_{\sigma \sigma \tilde{\cO}_{\text{new}_2}}$ can be calculated in the same manner as was done for $C_{\sigma \sigma \tilde{\cO}}$ above, the result is summarized as 
\begin{equation}
    \frac{C_{\sigma \sigma \tilde{\cO}_{\text{new}1}}^2 + C_{\sigma \sigma \tilde{\cO}_{\text{new}2}}^2}{C_{\sigma \sigma \tilde{\cO}}^2} = \frac{\sum_{i\neq j} (h_i^3 h_j -h_i^2 h_j^2) }{54 h_1 h_2 h_3}.
\end{equation}
This result should be invariant under how the two primaries were chosen.

In principle, this calculation can be extended to include higher-order descendants of the seed operators, but we will stop here. Splitting the result to only keep the holomorphic half, the result to the order calculated is then given by
\begin{equation}
    \begin{split}
        \cF^{\rY_{3,2}}(h_1, h_2, h_3; x) &= 3^{1-3(h_1+h_2+h_3)} \bigg(\cF_{\sigma\sigma}^{\sigma\sigma}(h_1+h_2+h_3; x) \\
        &\quad+ \frac{\sum_{i\neq j} (h_i^3 h_j -h_i^2 h_j^2) }{54 h_1 h_2 h_3} \cF_{\sigma\sigma}^{\sigma\sigma}(h_1+h_2+h_3+1; x) \bigg) + O\qty(x^{h_1 + h_2 + h_3 + 2 - 2h_\sigma}).
    \end{split}
\end{equation}
The genus two blocks in this frame were also computed in \cite{Cho:2017fzo}\footnote{The blocks $\cF_c(h_1, h_2, h_3|x)$ in \cite{Cho:2017fzo} are equal to $\frac{1}{3} \cF^{\rY_{3,2}}(h_1, h_2, h_3; x)$.} and the two calculations agree to the order computed here.

\section{Semiclassical limits}\label{a:semiclassics}
In this appendix, we derive the semiclassical limits of the functions used throughout the paper. We start with the DOZZ formula and reflection coefficients and move on to the conformal blocks for the sphere with four insertions and $\bbZ_3$-symmetric genus two surfaces.

\subsection{DOZZ formula and reflection coefficients}\label{a:semiclassics:dozz}
The DOZZ formula \cite{Dorn:1994xn,Zamolodchikov:1995aa} is the exact expression for the three-point structure coefficients,
$$C(\alpha_1,\alpha_2,\alpha_3) = \mel{\alpha_3}{V_{\alpha_2}(1,1)}{\alpha_1},$$
in Liouville theory, given by
\begin{equation}\label{eq:semiclassics:dozz}
    \begin{split}
    &C\qty(\alpha_1, \alpha_2, \alpha_3) = \qty[\pi \mu \gamma\qty(b^2) b^{2-2 b^2}]^{\frac{Q-\sum \alpha_i}{b}} \\
    &\quad\times \frac{\Upupsilon_0 \Upupsilon_b\qty(2 \alpha_1) \Upupsilon_b\qty(2 \alpha_2) \Upupsilon_b\qty(2 \alpha_3)}{\Upupsilon_b\qty(\alpha_1+\alpha_2+\alpha_3-Q) \Upupsilon_b\qty(\alpha_1+\alpha_2-\alpha_3) \Upupsilon_b\qty(\alpha_2+\alpha_3-\alpha_1) \Upupsilon_b\qty(\alpha_1+\alpha_3-\alpha_2)},
    \end{split}
\end{equation}
where $\Upupsilon_b(\alpha)$ is defined by
\begin{equation}
    \log \Upupsilon_b(\alpha)=\int_0^{\infty} \frac{\diff t}{t}\qty[\qty(\frac{Q}{2} - \alpha)^2 \ep^{-t}-\frac{\sinh ^2\qty(\qty(\frac{Q}{2} - \alpha) \frac{t}{2})}{\sinh \frac{t b}{2} \sinh \frac{t}{2 b}}], \qquad 0 < \Re \alpha < Q,
\end{equation}
for real and positive $b$. This integral representation is only defined for $0 < \Re \alpha < Q$ and has an analytic continuation in $\alpha$. In this expression, $\Upupsilon_0$ and $\gamma(\alpha)$ are
\begin{equation}
    \Upupsilon_0 = \eval{\frac{\diff \Upupsilon_b(\alpha)}{\diff \alpha}}_{\alpha = 0}, \qquad \gamma(\alpha) = \frac{\Gamma(\alpha)}{\Gamma(1-\alpha)}.
\end{equation}
The expression in equation~(\ref{eq:semiclassics:dozz}) obeys
\begin{equation}
    C\qty(Q-\alpha_1,\alpha_2,\alpha_3) = R\qty(\alpha_1) C\qty(\alpha_1, \alpha_2, \alpha_3),
\end{equation}
for
\begin{equation}
        R(\alpha) =\qty[\pi \mu \gamma\qty(b^2) b^{2-2 b^2}]^{\frac{2 \alpha-Q}{b}} \times \frac{\Upupsilon_b\qty(2 \alpha_1-Q)}{\Upupsilon\qty(2 \alpha_1)}.
\end{equation}

In the semiclassical limit, the classical action $S_3$ in equation~(\ref{eq:liouville:dozzexp}) for sphere with three singularities is
\begin{equation}
    \begin{split}
    S_3(p_1,p_2,p_3) &= - \lim_{b \to 0} b^2 \log C\qty(\alpha_1, \alpha_2, \alpha_3) \\
    &= \sum_{\sigma_i= \pm 1} F\qty(\frac{1}{2}+\ic p_1+\ic \sigma_2 p_2+\ic \sigma_3 p_3)+\sum_{j=1}^3\qty(H\qty(2 \ic p_j)+\pi\qty|p_j|) \\
    &\quad+ \ic \sum_{j=1}^3 2 p_j\qty(1-\log \qty|p_j|+\frac{1}{2} \log \qty(\pi \mu b^2)) + \text{const.},
    \end{split}
\end{equation}
where
\begin{equation}
    \lim_{b\to 0} b \alpha_i = \frac{1}{2} + \ic p_i,
\end{equation}
and the functions $F(x)$ and $H(x)$ are defined as
\begin{equation}
    F(x)=\int_{\frac{1}{2}}^x \diff y \, \log \frac{\Gamma(y)}{\Gamma(1-y)}, \qquad H(x)=\int_0^x \diff y \, \log \frac{\Gamma(-y)}{\Gamma(y)}.
\end{equation}
The contribution of the reflection coefficient $r(p)$ in equation~(\ref{eq:liouville:refexp}) is
\begin{equation}
    \begin{split}
        r(p) &= - \lim_{b\to 0} b^2 \log R(\alpha) \\
        &= - 4 \ic p\qty(1-\log |p|+\frac{1}{2} \log \qty(\pi \mu b^2)).
    \end{split}
\end{equation}

\subsection{Four-point conformal blocks}\label{a:semiclassics:4pt}
The four-point conformal blocks $\cF_{34}^{21}(h;q)$ can be computed to arbitrary order in $q$ (or equivalently cross-ratio $x$) using Zamolodchikov's recursive formula \cite{Zamolodchikov:1987}. In elliptic representation, we have
\begin{equation}
    \cF_{34}^{21}(h;q) = x^{\frac{Q^2}{4} - h_1 - h_2}(1-x)^{\frac{Q^2}{4} - h_2 - h_3} \Theta_3(\tau)^{3Q^2 - 4\sum_i h_i} (16q)^{h - \frac{Q^2}{4}} H_{34}^{21}(h;q),
\end{equation}
where the dimension $h_i$ is related to Liouville momentum $\alpha_i$ by $h_i = \alpha_i(Q-\alpha_i)$. In this expression, $q = \ep^{\ic \pi \tau}$ is the nome for the parameter $\tau$ defined in equation~(\ref{eq:torusparameter}) on the upper half-plane. The function $H_{34}^{21}(h;q)$ admits a power series around $q = 0$,
\begin{equation}
    H_{34}^{21}(h;q) = \sum_{n=0}^\infty H_n q^n,
\end{equation}
which can be computed using the recursion relations. In the semiclassical limit, the classical blocks are
\begin{equation}
    \begin{split}
        f_{34}^{21}(p; q) &= \lim_{b\to 0} b^2 \cF_{34}^{21}(h;q) \\
        &= p\text{-const.} + p^2 \log 16q + \lim_{b\to0} b^2 \log(H_{34}^{21}(h;q))
    \end{split}
\end{equation}
Since we are interested in the derivatives of the conformal blocks with respect to the exchange momentum, we only consider the $p$-dependent part of the classical blocks. For the case with identical external operators with Liouville momentum $p_\circ$, we find
\begin{equation}
    f_\circ(p;q) = p\text{-const.} + p^2 \log 16q + \frac{\qty(1+16p_\circ)^2}{4\qty(1+p^2)} q^2 + O\qty(q^4).
\end{equation}

\subsection{Genus two conformal blocks}\label{a:semiclassics:gen2}
The conformal blocks on $\rY_{3,2}$ are
\begin{equation}
    \begin{split}
        \cF^{\rY_{3,2}}(h_1,h_2,h_3;q) &= \qty(z(1-z))^{\frac{c}{8} - 2h_\sigma} \Theta_3(\tau)^{\frac{3c}{2} - 16h_\sigma} q^{-\frac{c}{8}} \\
        &\quad\times \qty(\frac{16q}{27})^{h_1 + h_2 + h_3} A^{\rY_{3,2}}(h_1,h_2,h_3;q),
    \end{split}
\end{equation}
where $h_\sigma = \frac{c}{9}$ is the dimension of twist three operator (see \S\ref{a:twistblocks}) and $A^{\rY_{3,2}}(h_1,h_2,h_3;q)$ can be written as a power series
\begin{equation}
    A^{\rY_{3,2}}(h_1,h_2,h_3;q) = \sum_{n=0}^\infty A_n q^n,
\end{equation}
whose coefficients are reported up to $O\qty(q^2)$ in \cite{Cho:2017fzo}. In the semiclassical limit, we find
\begin{equation}
    \begin{split}
        f^{\rY_{3,2}}(p_1,p_2,p_3;q) &= \lim_{b \to 0} b^2 \cF^{\rY_{3,2}}(h_1,h_2,h_3;q) \\
        &= p_i\text{-const.} + (p_1^2 + p_2^2 + p_3^2) \log \frac{16q}{27} \\
        &\quad+ \frac{8}{27}\qty(\frac{\qty(p_1^2-p_2^2)^2}{p_3^2+\frac{1}{4}}+\frac{\qty(p_1^2-p_3^2)^2}{p_2^2+\frac{1}{4}}+\frac{\qty(p_2^2-p_3^2)^2}{p_1^2+\frac{1}{4}}) q + O\qty(q^2).
    \end{split}
\end{equation}

\section{Numerical computation of genus two eigenvalues}\label{a:eigs}
We now examine the eigenvalue spectrum on the moduli space of $\bbZ_3$-symmetric genus two surfaces, denoted $\rY_{3,2}$. Having illustrated the computation of the metric in \S\ref{s:genus2Z3}, the eigenvalue equation is given by
\begin{equation}\label{eq:WPeig}
    \Delta^{\WP} \psi_n = E_n \psi_n.
\end{equation}
For simplicity, we will only study the case where the wavefunctions satisfy Dirichlet boundary conditions on the half fundamental domain in figure~\ref{fig:funDomainT}. On the moduli space $\rY_{3,2}$ this is equivalent to studying wavefunctions that are symmetric under the anharmonic group $\rS_3$ and antisymmetric under the $\bbZ_2$ corresponding to complex conjugation (note that in the $\tau$ coordinate this $\bbZ_2$ acts as $\tau \leftrightarrow -\conj{\tau}$). Explicitly the wavefunctions that we study satisfy the relations
\begin{equation}
    g \cdot \psi(\tau) = \psi(g^{-1} \cdot \tau) = \psi(\tau), \qquad \text{for} \; g \in \rS_3,
\end{equation}
and
\begin{equation}
    K \cdot \psi(\tau) = \psi(K \cdot \tau) = -\psi(\tau), \qquad \text{for} \; K \in \bbZ_2.
\end{equation}

The eigenvalue equation can be solved by discretizing the half fundamental domain into triangles and then numerically solving equation~\ref{eq:WPeig}. Table~\ref{tab:eigs} lists the smallest 24 eigenvalues found for this system. We leave further investigation of the eigensystem on $\rY_{3,2}$, such as the statistics of the eigenvalues, to future work.

\begin{table}[ht]
    \begin{center}
    \begin{tabular}{c c c c c c c c}
        \hline
        15.455 & 25.374 & 35.530 & 44.624 & 47.585 & 57.620 & 65.113 & 70.583 \\
        80.576 & 85.188 & 91.628 & 95.457 & 103.831 & 110.697 & 117.436 & 123.257 \\
        129.093 & 136.833 & 141.308 & 148.049 & 154.546 & 156.616 & 163.353 & 171.961\\  \hline
    \end{tabular}
    \end{center}
    \caption{First 24 eigenvalues of the Laplacian on $\rY_{3,2}$. The side length of each discrete triangle in the half fundamental domain is taken to be smaller than $10^{-2}$.}
    \label{tab:eigs}
\end{table}

\bibliographystyle{JHEP}
\bibliography{refs}
\end{document}